\documentclass[twocolumn, twocolappendix]{aastex631}

\usepackage{multirow}
\usepackage{natbib}
\usepackage{amsmath}

\newcommand{\Hb}{H$\beta$}
\newcommand{\Ha}{H$\alpha$}
\newcommand{\NII}{[N\,\textsc{ii}]}
\newcommand{\SII}{[S\,\textsc{ii}]}

\newcommand{\MgIb}{Mg\,\textsc{i}\,\textit{b}\,$\lambda\lambda$5167, 5173, 5184}
\newcommand{\CaHK}{Ca\,\textsc{H+K}\,$\lambda\lambda$3969, 3934}
\newcommand{\CaII}{Ca\,\textsc{ii}\,$\lambda\lambda$8498, 8542, 8662}
\newcommand{\OIII}{[O\,\textsc{iii}]}

\newcommand{\FeII}{Fe\,\textsc{ii}}

\newcommand{\ergs}{\text{erg s$^{-1}$}}
\newcommand{\ergscm}{\text{erg s$^{-1}$ cm$^{-2}$}}

\newcommand{\kms}{\text{km s$^{-1}$}}

\newcommand{\SDSSIV}{SDSS \uppercase\expandafter{\romannumeral4}}
\newcommand{\SDSSLEGACY}{SDSS \uppercase\expandafter{\romannumeral1}/\uppercase\expandafter{\romannumeral2}}

\hypersetup{
   colorlinks,
   linkcolor={blue!88!black!80},
   citecolor={blue!88!black!80},
   urlcolor={blue!88!black!80}}
   
\shorttitle{AGN-Host Spectral Decomposition}
\shortauthors{Ren W. et al.}

\graphicspath{{./}{figures/}}

\begin{document}

\title{Prior-Informed AGN-Host Spectral Decomposition Using PyQSOFit}

\correspondingauthor{Wenke Ren, Hengxiao Guo}
\email{rwk@mail.ustc.edu.cn (WKR), hengxiaoguo@gmail.com (HXG)}

\author[0000-0002-3742-6609]{Wenke Ren}
\affiliation{CAS Key Laboratory for Research in Galaxies and Cosmology, Department of Astronomy, University of Science and Technology of China, Hefei, Anhui 230026, China}
\affiliation{School of Astronomy and Space Science, University of Science and Technology of China, Hefei 230026, China}
\affiliation{Kavli Institute for the Physics and Mathematics of the Universe, The University of Tokyo, Kashiwa, Japan 277-8583 (Kavli IPMU, WPI)}

\author[0000-0001-8416-7059]{Hengxiao Guo} 
\affiliation{Shanghai Astronomical Observatory, Chinese Academy of Sciences, 80 Nandan Road, Shanghai 200030, China} 

\author[0000-0003-1659-7035]{Yue Shen}
\affiliation{Department of Astronomy, University of Illinois at Urbana-Champaign, 1002 West Green Street, Urbana, IL 61801, USA} 
\affiliation{National Center for Supercomputing Applications, 1205 West Clark Street,
Urbana, IL 61801, USA}

\author[0000-0002-4419-6434]{John D. Silverman}
\affiliation{Kavli Institute for the Physics and Mathematics of the Universe, The University of Tokyo, Kashiwa, Japan 277-8583 (Kavli IPMU, WPI)}
\affiliation{Department of Astronomy, School of Science, The University of Tokyo, 7-3-1 Hongo, Bunkyo, Tokyo 113-0033, Japan}
\affiliation{Center for Data-Driven Discovery, Kavli IPMU (WPI), UTIAS, The University of Tokyo, Kashiwa, Chiba 277-8583, Japan}
\affiliation{Center for Astrophysical Sciences, Department of Physics \& Astronomy, Johns Hopkins University, Baltimore, MD 21218, USA}

\author[0000-0001-9947-6911]{Colin J. Burke}
\affiliation{Department of Astronomy, Yale University, 266 Whitney Avenue, New Haven, CT 06511, USA}
\affiliation{Department of Astronomy, University of Illinois at Urbana-Champaign, 1002 West Green Street, Urbana, IL 61801, USA} 

\author[0000-0002-2052-6400]{Shu Wang}
\affiliation{Department of Physics \& Astronomy, Seoul National University, Seoul 08826, Republic of Korea}

\author[0000-0002-4419-6434]{Junxian Wang}
\affiliation{CAS Key Laboratory for Research in Galaxies and Cosmology, Department of Astronomy, University of Science and Technology of China, Hefei, Anhui 230026, China} 
\affiliation{School of Astronomy and Space Science, University of Science and Technology of China, Hefei 230026, China}

\begin{abstract}
We introduce an improved method for decomposing the emission of active galactic nuclei (AGN) and their host galaxies using templates from principal component analysis (PCA). This approach integrates prior information from PCA with a penalized pixel fitting mechanism which improves the precision and effectiveness of the decomposition process. Specifically, we have reduced the degeneracy and over-fitting in AGN-host decomposition, particularly for those with low signal-to-noise ratios (SNR), where traditional methods tend to fail. By applying our method to 76,565 SDSS Data Release 16 quasars with $z<0.8$, we achieve a success rate of $\approx$ 94\%, thus establishing the largest host-decomposed spectral catalog of quasars to date. Our fitting results consider the impact of the host galaxy on the overestimation of the AGN luminosity and black hole mass ($M_{\rm BH}$). Furthermore, we obtained stellar velocity dispersion ($\sigma_\star$) measurements for 4,137 quasars. The slope of the $M_{\rm BH}-\sigma_\star$ relation in this subsample is generally consistent with previous quasar studies beyond the local universe. Our method provides a robust and efficient approach to disentangle the AGN and host galaxy components across a wide range of SNRs and redshifts.

\end{abstract}

\keywords{AGN host galaxies (2017), Astronomy data analysis (1858), Active galactic nuclei (16), Quasars (1319)}


\section{Introduction} \label{sec:intro}

The optical spectrum of active galactic nuclei (AGNs) is a powerful tool for probing the inner regions around supermassive black holes (SMBH), such as dust-attenuating structures, the broad-line region, and the accretion disk \citep{Gaskell2007, Guo2022, Fries2023}. It also contains rich information about the environment around SMBHs, including gas-phase metallicity, geometry, and dynamics. Due to the blending of the AGN spectrum with that of the host galaxy during spectroscopic observations, accurate spectral decomposition is crucial for both AGN and host sciences \citep[e.g.,][]{Bentz2009a}. 

As pointed out by previous studies, the host galaxy emission accounts for a notable fraction of the total flux, e.g., more than 30\% at rest frame 5100\AA\ on average, based on both image and spectral decomposition for low-redshift quasars \citep[luminous AGNs;][]{Shen2015a, Matsuoka2015, Li2021}. Non-removal of host galaxy emission can lead to an overestimation of the quasar continuum luminosity and thus black hole mass, thereby biasing the estimation of the AGN luminosity and mass functions \citep{Hopkins2007, Singh2014, Shen2012, Kelly2013}. Such overestimation may also skew the $M_{\rm BH}-\sigma_\star$ and the $M_{\rm BH}-M_\star$ relationships, particularly when compared to local relations \citep{Reines2015, Shen2015a}. Furthermore, mixing host emission into the quasar component can result in an underestimation of the quasar variability and emission line equivalent width, potentially leading to misleading conclusions in studies of AGN broad-line structures and dynamics \citep{Villforth2012, Varisco2018, Ren2023}. 

On the other hand, a high fractional host galaxy contribution offers opportunities to study the host galaxy properties and delve deeper into AGN-host coevolution \citep[e.g.,][]{Kauffmann2003a,Kormendy2013}. With decomposed spectra, studies can be conducted, including an examination of the $M_{\rm BH}-\sigma_\star$ relationship \citep[e.g.,][]{Shen2019}, analyses of the connections between AGN accretion rates and host galaxy star formation rates \citep[e.g.,][]{Jin2018}, and exploring the luminosity correlation between AGN and their host \citep[e.g.,][]{Jalan2023}. Both existing and forthcoming spectral surveys are expected to greatly expand AGN and quasar samples thus facilitating a more thorough statistical analysis of AGN-host coevolution.

However, precisely disentangling the AGN and host galaxy components from composite spectra is a challenging task. Both AGN and galaxy exhibit diverse spectral features, lacking definitive indicators to determine their relative contributions. For the host galaxy component, the spectral shape depends on the stellar population and star formation history, both of which vary widely across different AGN populations \citep[e.g.,][]{Kauffmann2003a,Silverman2009, Ni2023}. Similarly, the AGN continuum shape can vary significantly, not only among different objects but also within the same object over years \citep[e.g.,][]{Giveon1999,Webb2000,VandenBerk2004}. The presence of overlapping emission and/or absorption lines from both AGN and the host galaxy further adds to the complexity of the decomposition. Therefore, further developments of robust and efficient methods of spectral decomposition remain an important and significant challenge.

Several notable studies have attempted to separate host galaxy emission from AGN spectra. These methodologies can be broadly classified into three categories: physical multi-component decomposition, principal component analysis (PCA) decomposition using a 1D spectrum, and techniques that leverage supplementary data beyond a single-epoch (SE) 1D spectrum. Incorporating spatial information beyond 1D spectrum proves both feasible and effective, especially when dealing with a small sample of targets. For instance, broad-band high-resolution images are often used to determine the host fraction and correct the AGN's monochromatic luminosity measured from spectra \citep[e.g.,][]{Wang2014,Bentz2015}. \citet{Jahnke2007} distinguishes the AGN emission from its host by using 2D spectra to model the distinct spatial distributions of the point-like nucleus and the extended host galaxy. Additionally, both on- and off-axis spectra and integral field spectroscopy observations can be employed for this decomposition \citep{Wold2010,Mezcua2020,Riffel2023}.

Unlike methods that require additional data, the second category relies solely on 1D spectra and intuitively employs physical models for decomposition. For the AGN component, the power-law continuum, \FeII\ templates, and Gaussian emission line profiles are usually adopted. Shifted and broadened stellar templates are commonly utilized for modeling the host galaxy. However, due to the complexity and time-consuming nature of addressing non-linear fitting problems, studies focusing on host galaxy properties often simplify the modeling by representing the AGN component with a power-law \citep{Canalizo2012,Jin2018}. In contrast, studies that emphasize detailed AGN properties may simplify the host galaxy templates \citep{Calderone2017,Lu2022}.  A few groups employ both sophisticated AGN and host templates through step-by-step fitting \citep{Matsuoka2015,Oio2019} or Markov Chain Monte Carlo (MCMC) techniques \citep{Sexton2021} to avoid local minima and diverging solutions. These meticulous fitting approaches allow for the simultaneous measurement of both host galaxy and AGN properties. Despite the sophistication of these models, they typically require spectra with high signal-to-noise ratios (SNR), generally above 15 per pixel, which far exceeds the typical survey level of about 5 for the Sloan Digital Sky Survey (SDSS) quasars at $z$ = 0.5. Moreover, the implementation of MCMC requires extended computational time to achieve convergence. As a result, applying these methods broadly across larger datasets often poses significant challenges.

Included in the above list, the PCA method is an effective technique for spectral decomposition for large survey samples \citep[e.g.,][]{Greene2004, Greene2005a, VandenBerk2006, Shen2008a, Sun2015, Shen2015a}. It employs two sets of truncated PCA templates, one derived from quasar samples and the other from galaxy samples, to linearly determine their optimal fitting combination. By leveraging PCA templates that synthesize the traits of various emissions into a few number of representative eigenspectra \citep{Yip2004, Yip2004a, Li2005}, this method avoids dependence on specific local features, making it exceptionally suitable for bulk fitting tasks. Consequently, it achieves consistent performance across wide distributions of redshifts and spectral signal-to-noise ratios (SNR). Despite these advantages, existing implementations have not been fully successful since only a fraction of the sample can be successfully decomposed with this PCA method \citep[e.g.,][]{Rakshit2020}. A primary limitation is that PCA decomposition is not a physically-driven method; its scaling parameters lack direct physical interpretations, leading occasionally to unphysical model outputs.

The technical reason for the failed outcomes primarily stems from the combination of two PCA templates. The orthogonality between template spectra from different sets is not guaranteed. With compromised independence, the fitting can easily be influenced by model degeneracies, overfitting to spectral noise, and dust extinction. Thus, the best-fit models often contain unrealistic features. To address the overfitting and degeneracy issues, we propose a method that incorporates prior information on the eigencoefficient distribution. This approach ensures the dominance of the first eigenspectra in its corresponding model, effectively preventing ill-fitting results from the disproportionate influence of high-order eigenspectra. Such refinement offers a more reasonable decomposition for continuum measurements and subsequent fittings. We have incorporated this practice as a module within the existing quasar spectral fitting software {\tt PyQSOFit} \citep{2018ascl.soft09008G}, and provide a detailed description in \S\ref{sec:method}.

By applying this method to the low-redshift (z<0.8) subset of SDSS Data Release 16 (SDSS DR16) quasars, we have compiled an unprecedentedly large catalog of host-decomposed quasar spectra. Utilizing the host-free SE black hole mass estimation recipe \citep{Shen2023a}, we explored the impact of host galaxy contamination on the accuracy of measurements of AGN properties. We further examined how spectral SNR affects the reliability of the measurements and assessed the resulting $M_{\rm BH}-\sigma_\star$ relation.

This paper is organized as follows. The algorithm is described in \S\ref{sec:method}. We present the limitations of the linear decomposition method in \S\ref{subsec:limitation}. The detailed mechanism and the benchmark tests are described in \S\ref{subsec:PCA} and \S\ref{subsec:benchmark}. In \S\ref{sec:apply}, we apply our method to the SDSS DR16 quasar \citep[SDSS DR16Q,][]{Lyke2020a} catalog and present the refined results. We summarize our results in \S\ref{sec:summary}. Throughout this paper, we adopt a flat $\rm{\Lambda}$CDM cosmology with $\Omega_\Lambda=0.7$, $\Omega_m=0.3$, and $H_0=70~\rm{km~s^{-1}~Mpc^{-1}}$.

\section{Algorithm} \label{sec:method}

In the following subsections, the limitations and drawbacks of the existing linear PCA decomposition approach are initially discussed. This is followed by a detailed description of the fitting procedure for our proposed prior-informed PCA decomposition method. Subsequent sections present benchmark tests that evaluate the accuracy and efficiency of our method. The discussion and the benchmark tests are all based on the SDSS spectra unless otherwise specified. Note that our program can also be extended to other spectroscopic surveys, such as the Dark Energy Spectroscopic Instrument (DESI) \citep{DESICollaboration2016,DESICollaboration2016a} and the Subaru Prime Focus Spectrograph (PFS) \citep{Takada2014}. However, compatibility and performance may vary due to the differences in spectral resolution and sample selection.

\subsection{Limitations of Linear PCA Decomposition} \label{subsec:limitation}

\begin{figure*}[tb!]
   \includegraphics[width=1\textwidth]{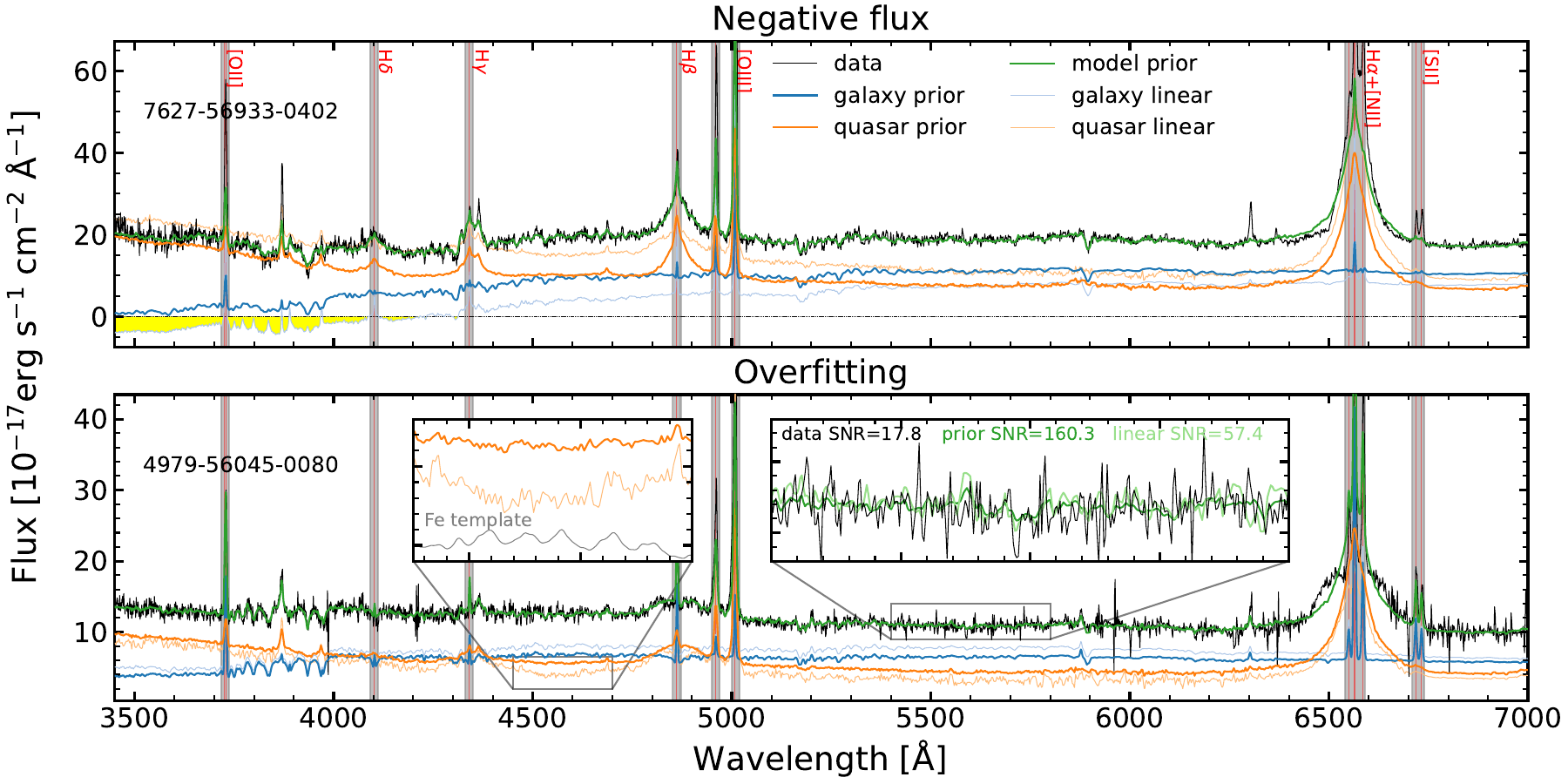}
   \caption{Two examples show typical problems (negative flux and overfitting) in the linear decomposition method. The first example highlights the negative flux of the host galaxy model (yellow regions) in linear decomposition. The second example shows the unrealistic twist and overfitting problem, indicated in insets. Both issues are addressed with our new method; see \S\ref{subsec:limitation} for details. The original spectra are depicted with black lines, while regions with strong emission lines, shaded in gray, are masked during fitting. Our fitted models for the galaxy and quasar are represented in dark blue and orange lines, respectively, with the combined model displayed in green. For comparison, the results from linear decomposition are shown in light blue and orange. Spectral names (plate, MJD, fiber), line names, SNR are listed.
   \label{fig:decomp_QA}}
\end{figure*}

The PCA, also known as the Karhumen-Lo\`{e}ve transformation, is a mathematical method usually used for dimensionality reduction, transforming high-dimensional data into a lower-dimensional space. The fundamental concept is to find a lower-dimensional set of orthogonal eigenvectors where the projection of the original data can preserve most of its characteristics. In spectral analysis, each spectrum can be thought of as a vector in a multidimensional hyperspace, with each wavelength pixel representing one axis, $\hat{\lambda}_{j}$. Then the $i$th spectrum can be expressed as $\mathbf{f}_{i}=\sum_{j}{f_{ij}\hat{\lambda}_{j}}$, where $f_{ij}$ is the spectral flux at $\hat{\lambda}_{j}$. Similarly, we can define a set of rotated base vectors, $\mathbf{e}_{k} = \sum_{j} e_{jk} \hat{\lambda}_{j}$, where $k$ is the index of the base vector. Thus, the original spectra can also be expressed as $\mathbf{f}_{i}=\sum_{k}{f'_{ik}}\mathbf{e}_{k}$, where $f'_{ik}=\sum_{j}{f_{ij} e_{jk}}$. Given a sample of spectra, the PCA aims to identify a set of base vectors that maximize the variance of the projections of the original data onto these vectors in a sequential manner: the first base vector captures the highest variance, denoted by $\sigma^2_{f'_{i1}}$, followed by the second base vector, which accounts for the next highest variance, and so forth. We call this set of basis vectors as the eigenspectra. Since the eigenspectra are usually built based on populations with similar characteristics, ideally, the first few eigenspectra would account for nearly all information (variance) \citep{VandenBerk2006}.

In astronomy, the PCA technique is extensively used for spectral classification and dimensionality reduction \citep[e.g.,][]{Connolly1995, Ronen1999, McGurk2010, Paris2011, Ma2019}. \citet{Yip2004, Yip2004a} and \citet{Li2005} have demonstrated that, for galaxies, the first five eigen spectra can account for over 98\% of the variance, while for quasars, the first ten eigenspectra can explain more than 90\% of the variance. If AGN spectra are simple combinations of emissions from quasars and galaxies, integrating the PCA spectra of both allows us to effectively separate two components from the composite spectra \citep{VandenBerk2006, Shen2008a}. Thus, the method of decomposing using a linear combination of PCA spectra has been widely adopted in various studies due to its computational efficiency and stability \citep[e.g.,][]{Shen2015a, Sun2015, Rakshit2020}.

However, when these methods are applied to a large quasar sample with relatively low spectral SNR of $\sim$ 5, typical of current spectroscopic surveys, the failure rate of such decompositions increases significantly \citep{Rakshit2020}. Moreover, even for high SNR spectra, there remains a substantial failure rate. We emphasize that these failures stem not only from extremely low host galaxy fractions, but also from significant contributions of model degeneracies and overfitting.

Fig. \ref{fig:decomp_QA} exhibits two typical examples of the unreliable decomposition results using 10 quasar eigenspectra and 5 galaxy eigenspectra. Negative model is the most common problem that leads to failure in linear decomposition, as shown with at the shorter wavelength range in the upper panel (also see Fig. \ref{fig:efficacy}). Even for some cases where the decomposed models are entirely positive, they might still be physically unreal solution, e.g., absorption features erroneously appearing in regions typically characterized by emission lines. These issues are often caused by competition between two sets of PCA templates. If one set overestimates the flux, the other must compensate by turning negative to offset the excess and align closely with the total data flux. 

In the second example, we clearly see a depressed feature in the linearly decomposed quasar model at around 4600 \AA\ (also see the inset), where the original data are smooth and flat. Normally, this region is dominated by a series of broadened \FeII\ emission lines, and thus, a flat or convex continuum would be expected. We suggest that the concave decomposed AGN continuum is anomalous, likely resulting from degeneracies. Furthermore, the linear decomposition method tends to overfit the noise within the spectra. Theoretically, the fitted template should maintain a constant SNR level, independent of the data's SNR. In the second example the linearly decomposed AGN model (light green line) appears significantly noisier than the prior-informed model (dark green line) around 5600 \AA\ (also see the inset). Given the high SNR of the PCA templates, there are no prominent features in this region, and therefore, the fitted model should exhibit a very smooth appearance. The additional noise in the linearly fitted results stems from overfitting the data noise with high-order PCA templates (see \S\ref{subsec:benchmark} for details).

The issues mentioned above are primarily due to the simplistic nature of the decomposition method. A critical problem is that although the PCA spectra within the same set are mutually orthogonal and independent, those from two different sets (i.e., quasar and galaxy template libraries) can exhibit severe degeneracy. Furthermore, in cases of low SNR spectra, high-order PCA templates might be used to fit noise artifacts, resulting in their disproportionate weighting in the fitting process (see \S\ref{subsubsec:prior}). Consequently, originally negligible higher-order eigenspectra might compete for flux with the dominant eigenspectra, leading to a disastrously unrealistic solution.

\subsection{Prior-Informed PCA Decomposition} \label{subsec:PCA}

To mitigate issues of degeneracy and overfitting, we assign weights to each PCA eigenspectra during the decomposition. These weights are natural byproducts of the construction of PCA templates. We employ the penalized pixel fitting method \citep[e.g.,][]{Merritt1997} to incorporate these weights into the fitting procedure. By leveraging this approach, we effectively mitigate the degeneracy between different PCA templates and significantly enhance the overall quality of the decomposition. The success rate of this method reaches approximately 94\% for the SDSS DR16Q catalog with $z<0.8$.

\subsubsection{Distribution of PCA Eigencoefficients} \label{subsubsec:prior}

Following \citet{VandenBerk2006}, we chose the quasar PCA templates built based on the C1 bin\footnote{To ensure adequate spectral coverage and property similarity, \citet{Yip2004} divided their full sample into four luminosity ranges ($M_i=-22$ to $-30$, labeled A to D) and five redshift ranges (ZBIN 1 to 5 spanning $z=0.08$ to $5.13$). The subsample we used occupies the most luminous 6\% within the lowest redshift range.} in \citet{Yip2004}, which has an absolute $i$-band magnitude of $-26<M_i<-24$ and a redshift of $0.08<z<0.53$. The low redshift range was selected to match the general population properties of our target sample. The high luminosity of the PCA templates minimizes the host contamination. However, since these PCA templates were constructed based on the first data release (DR1) of SDSS and include a relatively small number of identified quasars (109 objects), they may not fully represent the broader quasar population. While the current templates are sufficient for proof of concept, developing more comprehensive PCA templates \citep{Bailey2012}, e.g., $z$ up to 1, in a future study could further improve the feasibility and accuracy of the host decomposition \citep{Brodzeller2023}. The galaxy PCA templates used in this study were created by \citet{Yip2004a} and are based on approximately 170,000 galaxies from SDSS DR1.

\begin{figure*}[tb!]
   \includegraphics[width=1\textwidth]{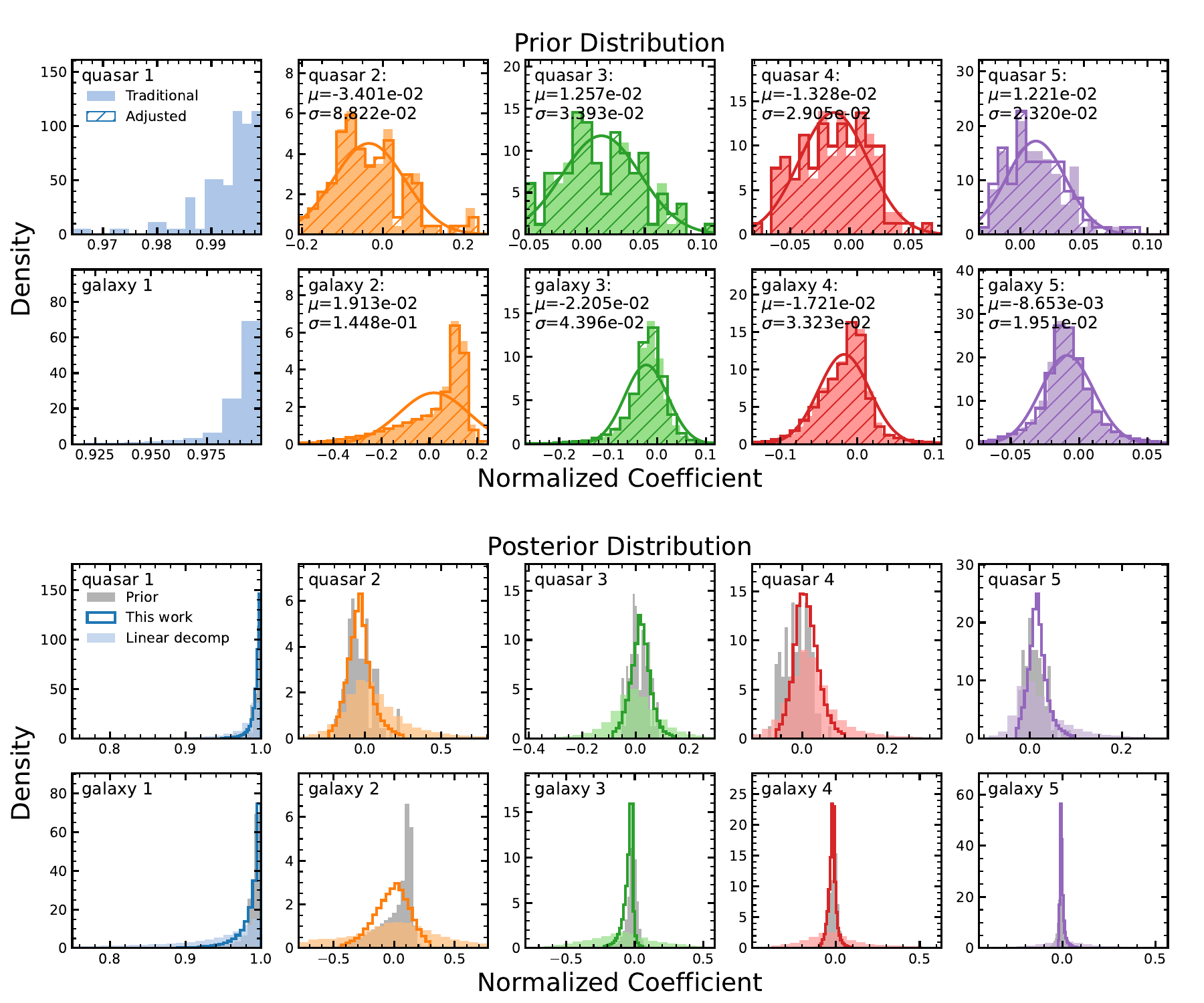}
   \caption{
      The prior and posterior distributions of the eigencoefficients.
      The upper panels show the distributions of normalized eigencoefficients for the first five quasar and galaxy PCA templates. The filled (traditional) and dashed (adjusted) histogram represent the distribution of coefficient normalized with different methods (See \S\ref{subsubsec:ppxf}). Our adjusted method can faithfully recover the traditional distribution and enhance the optimization efficiency. We use one Gaussian to model the adjusted distribution, and the fitted parameters are listed in the upper left. The lower panels show the posterior distributions of eigencoeffcients from prior-informed and the distributions from linear decomposition based on SDSS DR16 quasars. Prior distributions are plotted for comparison. The prior-informed method significantly reduced the fitting degeneracy and the scatter of the eigencoeffcients distribution.
   \label{fig:prior}}
\end{figure*}

Following the procedure described in \citet{Yip2004, Yip2004a}, we collected the spectra used to construct the PCA templates and then projected the data onto each eigenspectrum to obtain the coefficient,
\begin{equation}
c_i=\sum_{k}{f_{k}e_{ik}},
\end{equation}
where $f_{k}$ is the spectral flux at wavelength $\lambda_k$ and $e_{ik}$ is the $i$th eigenspectra value at wavelength $\lambda_k$. We retained only the coefficients associated with the first ten eigenspectra and normalized these coefficients for each spectrum to ensure that the sum of their squares equals 1. These normalized coefficients, known as eigencoefficients or weights, are illustrated in the upper panels of Fig. \ref{fig:prior}. The distributions of the first five eigencoefficients for both the quasar and galaxy samples are presented. 

For galaxy and quasar spectra, the first eigenspectrum accounts for over 90\% of the total weight, while the second eigenspectrum takes about half of the remaining weight, on average. Assuming that the AGN spectra comprise two independent components (quasar$+$galaxy), we would expect that the normalized eigencoefficients for each component derived from the decomposition process should generally follow the distributions observed in pure quasar and galaxy populations respectively. However, in practice, the linear decomposition method does not guarantee these expected distributions.

In the lower panels of Fig. \ref{fig:prior}, we apply linear decomposition to quasars from the SDSS DR16Q catalog and present the distribution of the eigencoefficients as color-filled histograms. For comparison, we also display the posterior distribution obtained through our prior-informed method as step histograms, alongside the prior distribution shown in gray histograms. The scatter in the coefficient distribution derived from the linear method is significantly larger than expected, suggesting that this method has overestimated the weights of high-order PCA templates in a substantial portion of the sample. Consequently, we propose using the parameter distribution from the building sample as a prior to constrain the fitting. As illustrated in the figure, our prior-informed method effectively reduces the scatter and confines the parameter distribution within a reasonable range.

In the linear decomposition method, we notice a greater deviation in the parameters for galaxy PCA templates compared to those for quasar PCA templates. This disparity arises because the quasar component typically dominates the total flux, resulting in more stable fitting outcomes. Conversely, the galaxy component is relatively weaker and more sensitive to noise, resulting in more serious fitting degeneracy.

\subsubsection{Penalized Pixel Fitting} \label{subsubsec:ppxf}

The fundamental reason for the mismatch between the linear PCA decomposed distributions and the prior distribution stems not only from the degeneracy among PCA templates but also from the overfitting of noise. To resolve this issue, it is crucial to impose self-adaptable constraints on the fitted PCA eigencoefficients, which can change the strength of the constraints according to the spectral SNR. The maximum penalized likelihood formalism used in {\tt pPXF} provides a solution to this requirement \citep{Merritt1997, Cappellari2004, Cappellari2016, Cappellari2023}. The basic idea is to incorporate an adjustable penalty term to the $\chi^2$ to move the solution towards the prior. The penalty term is only related to the weights of each fitting parameter. The greater the deviation of the weights from the prior, the larger the penalty term becomes. Since the $\chi^2$ term is also affected by the spectral SNR, the penalty term would be more dominant in lower SNR cases. By combining these two terms within a balance factor, the program can automatically make a compromise between a lower $\chi^2$ and a more prior-like solution.

To construct a uniform penalty term that incorporates both the galaxy and quasar PCA sets, it is necessary to normalize each set of coefficients. Traditionally, this normalization involves:
\begin{equation}
q\prime_i=\frac{q_i}{\sqrt{\sum_{i=1}^{m}{q_i^2}}},
\end{equation}
where $q_i$ and $q\prime_i$ represent the eigencoefficients before and after normalization, respectively, and $m$ denotes the total number of templates used. This normalization method is equivalent to setting an implicit constraint to all coefficients, i.e. the sum of the squares of all coefficients equals 1, which would hinder the optimization. Considering the computational costs and convergence efficiency, we use the coefficient of the first eigenspectra as the normalization factor and only use priors to constrain the rest of the coefficients. The normalization formula is then simplified to:
\begin{equation}
q\prime_i=\frac{q_i}{q_1}, i\neq 1.
\end{equation}
The new distributions of the rest of the coefficients are shown as hatched histograms in Fig. \ref{fig:prior}. Since the first eigencoefficients in both sets consistently approximate 0.95 with little variation, re-normalizing the subsequent coefficients does not substantially alter the prior distribution.

Given normalized eigencoefficients, we can then parameterize the prior distribution. Except for the first-order eigencoefficients, the rest generally distribute around 0 with a small standard deviation. We thus simply use a normal distribution to represent each prior distribution. The parameter used in this work is listed in Table \ref{tab:prior_pp}. Apparently, the distributions of the second-order galaxy eigencoefficients are asymmetrical and have a longer tail than the normal distribution. However, the approximation using a normal distribution is sufficiently effective for improving decomposition fitting. Given that the penalty term must be calculated in every iteration of the optimization, employing a complex distribution would substantially increase the computational cost with negligible gain. Additionally, the data size available for quasar PCA templates is not large enough to support a more intricate model. Constructing quasar PCA templates with a data volume comparable to that of galaxy PCA templates, and modeling the distribution using GAMLSS \citep[Generalized Additive Models for Location, Scale, and Shape,][]{Rigby2005} or a non-parametric model, would be a promising future work.

With the above preparations in place, we design the penalized $\chi_p^2$ as
\begin{equation}
   \chi_p^2=\chi^2 (1 + \alpha \mathcal{P}),
\end{equation}
where $\alpha$ is the adjustable factor and $\mathcal{P}$ is the penalty function, respectively. Assuming every normalized coefficient $q\prime_i=q_i/q_1~(i\neq 1)$ follows a normal distribution, we can fit a mean value $\mu_{\rm q\prime_{i}}$ and a standard deviation $\sigma_{\rm q\prime_{i}}$ from the prior sample. Then, the penalty function can be naturally constructed as:
\begin{equation}
\mathcal{P}=\left(\sum_{i=2}^{m}{\left(\frac{q\prime_{i}-\mu_{q\prime_{i}}}{\sigma_{q\prime_{i}}}\right)^2}+\sum_{i=2}^{n}{\left(\frac{g\prime_{i}-\mu_{g\prime_{i}}}{\sigma_{g\prime_{i}}}\right)^2}\right)/(m+n-2).
\end{equation}
Here, we use $q_i$ ($g_i$) to represent the $i$th coefficient of quasar (galaxy) PCA template and $m$ ($n$) to represent the total number of used template. 

During our fitting, we adopted the penalty factor $\alpha=0.01$, which has the most prior-like posterior distribution. With a larger penalty factor, the fitting procedure would have stronger constraints and resulted in a narrower posterior distribution of the fitted eigencoefficients and vice versa. Nevertheless, the success rate of decomposition and the host fraction would not be significantly affected by changing $\alpha$ from 0.001 to 0.1. 

Throughout this work, we use the first 5 PCA components for galaxies and the first 10 PCA components for quasars to perform the decomposition, i.e., $m=10$ and $n=5$, following previous studies \citep[e.g.,][]{VandenBerk2006, Shen2008a, Shen2015a, Rakshit2020}. As all PCA eigencoefficients are constrained by the prior distributions, incorporating additional higher-order PCA components, which contribute minimally to the overall weight, would not substantially alter our results. Major emission lines are masked during the fitting process to mitigate the model competition effect (gray windows in Fig. \ref{fig:decomp_QA}). We eventually reconstructed the quasar and galaxy eigenspectra, utilizing {\tt LMfit} \citep{Newville2014} for minimization, and measured the host galaxy fraction at 5100 \AA\ ($f_{\rm h,5100}$) and 4200 \AA\ ($f_{\rm h,4200}$).

\begin{deluxetable}{cCCCC}
   \tablecaption{Best-fit parameters of the normal distribution for the first 10 normalized PCA eigencoefficients of quasar and galaxy sets. \label{tab:prior_pp}}
   \tablecolumns{5}
   \tablewidth{0pt}
   \tablehead{
   \colhead{Order} & \colhead{$\mu_{\rm gal}$} & \colhead{$\sigma_{\rm gal}$} & \colhead{$\mu_{\rm qso}$} & \colhead{$\sigma_{\rm qso}$}
   }
   \startdata
   1 & /        & /        & /        & /        \\
   2 & $~~1.91\times10^{-2}$ & $1.45\times10^{-1}$ & $-3.40\times10^{-2}$ & $8.82\times10^{-2}$ \\
   3 & $-2.20\times10^{-2}$ & $4.40\times10^{-2}$ & $~~1.26\times10^{-2}$ & $3.39\times10^{-2}$ \\
   4 & $-1.72\times10^{-2}$ & $3.32\times10^{-2}$ & $-1.33\times10^{-2}$ & $2.90\times10^{-2}$ \\
   5 & $-8.65\times10^{-3}$ & $1.95\times10^{-2}$ & $~~1.22\times10^{-2}$ & $2.32\times10^{-2}$ \\
   6 & $-1.20\times10^{-2}$ & $2.12\times10^{-2}$ & $~~4.47\times10^{-3}$ & $1.74\times10^{-2}$ \\
   7 & $~~5.63\times10^{-3}$ & $1.67\times10^{-2}$ & $-1.36\times10^{-3}$ & $2.03\times10^{-2}$ \\
   8 & $-4.41\times10^{-3}$ & $1.26\times10^{-2}$ & $-1.36\times10^{-3}$ & $1.16\times10^{-2}$ \\
   9 & $-3.45\times10^{-3}$ & $9.84\times10^{-3}$ & $-5.25\times10^{-3}$ & $6.17\times10^{-3}$ \\
   10 & $~~4.85\times10^{-3}$ & $7.82\times10^{-3}$ & $~~4.21\times10^{-3}$ & $9.57\times10^{-3}$ \\
\enddata
\end{deluxetable}

\subsection{Benchmark Tests} \label{subsec:benchmark}

\subsubsection{Efficiency} \label{subsubsec:efficiency}

Utilizing a comprehensive stellar library as the host template for performing host spectral decomposition is a relatively robust approach. This method fits all components simultaneously and employs an MCMC technique to ascertain the global optimum \citep[e.g.,][]{Sexton2021}. However, the extensive computational demands of the MCMC method, which typically require several hours per spectrum, making it impractical for large-scale studies that involve tens of thousands of quasar spectra.

The prior-informed decomposition proposed by this paper is more economical. The entire fitting procedure, combining the decomposition process, host galaxy fitting with {\tt pPXF}, and quasar spectra fitting with {\tt PyQSOFit}, only requires a few minutes for a single computational core to execute 100 Monte Carlo simulations to assess statistical uncertainties. To avoid convergence at a local minimum, we run decomposition with initial host fractions from 0.1 to 0.9 in increments of 0.1 for each spectrum and adopt the result with the lowest $\chi^2$. The overall resource consumption of our pipeline approximately triples that of the original {\tt PyQSOFit}, which adopts the linear combination decomposition method.

\subsubsection{Efficacy} \label{subsubsec:efficacy}
\begin{figure}[tb!]
   \includegraphics[width=0.48\textwidth]{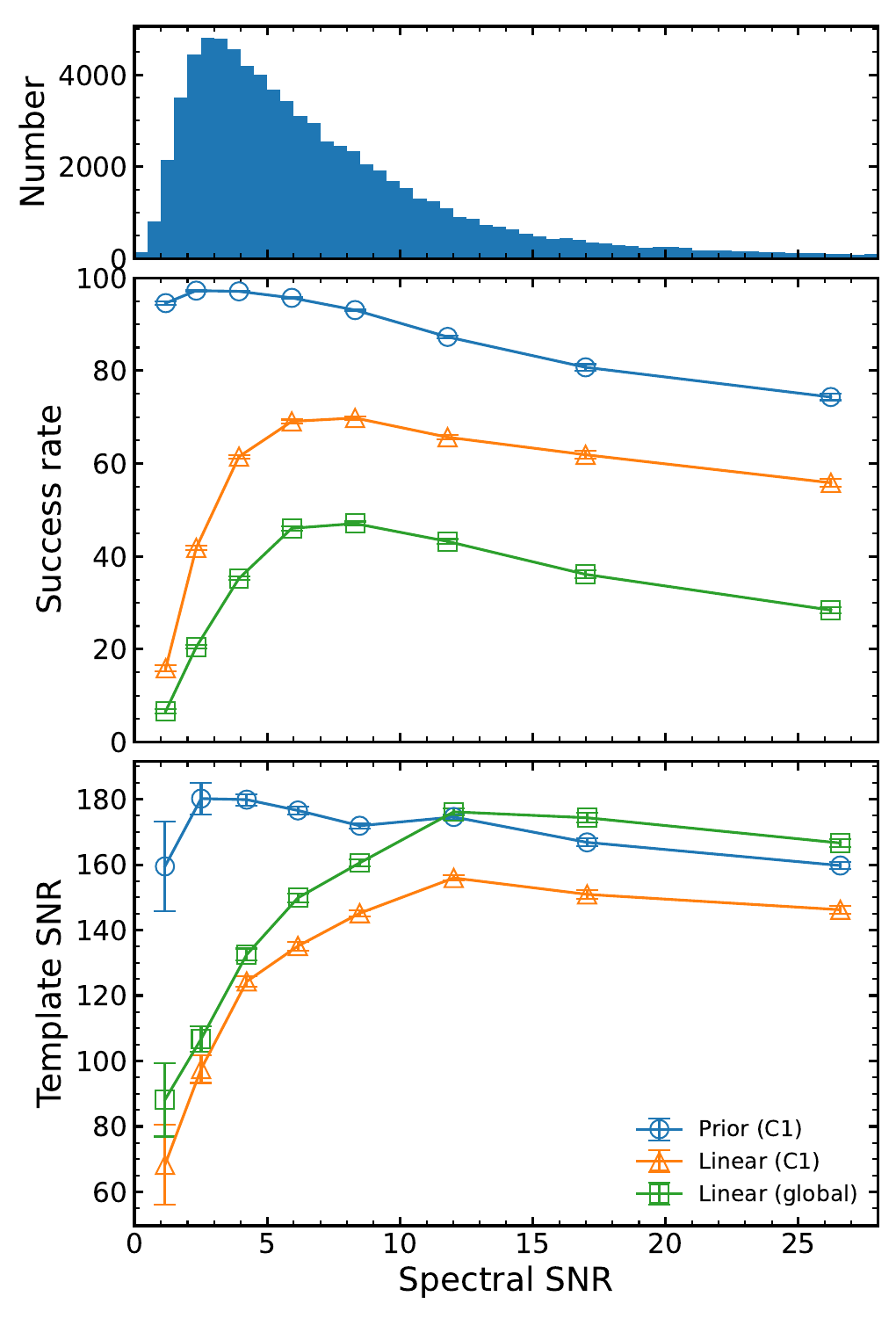}
   \caption{Upper panel: The number density distribution of original sample used in decomposition. Middle panel: The success rate of different decomposition methods as a function of spectral SNR. Blue circles represent our prior-informed method; orange triangles stand for the linear decomposition method using quasar PCA templates based on the C1 sample (a low redshift and high luminosity quasar sample) in \citet{Yip2004}; green squares denote the linear decomposition method using global quasar PCA templates. Lower panel: the mean SNR of reconstructed model of all successfully decomposed spectra as a function of spectral SNR. If the model is not overfitting to spectral noise, a uniformly high spectral SNR of the model is expected.
   \label{fig:efficacy}}
\end{figure}

The decomposition success rate and the ability to avoid overfitting noise are two key evaluation metrics for our method. To assess the efficacy of our approach, we applied our decomposition process to the low-redshift subset of SDSS DR16 quasars, which comprise a total of 76,565 quasars with $z<0.8$ (see \S\ref{subsec:data} for details).

Given the definition of failed host decomposition, where the fitted galaxy or quasar templates exhibit more than 10\% negative pixels, we assessed the efficacy of our prior-informed method across different spectral SNRs, compared with that of linear decomposition methods, as shown in Fig. \ref{fig:efficacy}. Two different templates are used in the linear decomposition method: the C1 bin and global quasar PCA templates. As previously mentioned, the former is built from a low-redshift, high-luminosity subsample, adopted in our developed decomposition method. The latter is the default template used in previous versions of {\tt PyQSOFit}, and was constructed from all quasar spectra in SDSS DR1, covering a redshift range up to $\sim$ 5 and a wavelength span from 900 to 8000 \AA. The spectral SNRs are calculated using a data-based algorithm, {\tt DER\_SNR}\footnote{{\tt DER\_SNR} uses only the flux data to assess the SNR. It takes the median flux of the spectrum as the signal and the median absolute deviation of flux over all pixels as the noise. The absolute deviation of each pixel is calculated locally with its neighboring pixels.} \citep{Stoehr2008}, which provides a simple and robust estimation for spectra with varying wavelength coverage, without relying on pixel errors.

The middle panel of Figure \ref{fig:efficacy} illustrates two significant improvements achieved through our prior-informed decomposition approach: Firstly, the success rate of our method has significantly improved across all spectral SNR regimes, outperforming the linear decomposition. Secondly, by incorporating a penalty function and C1 quasar PCA templates, we effectively mitigate the degeneracy issue in most SDSS quasars, particularly at the low SNR end (less than 5). This has elevated the success rate from $<$ 60\% to $>$ 90\%. 

The difference of using different PCA templates is clearly demonstrated in two distinct linear decompositions (C1 and global). Although global quasar PCA templates are generally considered more representative of typical quasars, the inclusion of low-luminosity quasars in their constructed sample introduces potential host features. These features compete with the galaxy PCA template, leading to failed decompositions. Additionally, during the decomposition process, a significant portion of the global PCA template must be truncated, further exacerbating the degeneracy issue. Consequently, it is not surprising that the success rate of the global PCA templates is lower than that of the C1 bin templates.

For the prior-based results, the success rate decreases with increasing spectral SNR. This occurs because high-SNR spectra retain finer structures, necessitating either a greater number of templates or a more intricate model for accurate fitting. While increasing the number of PCA templates used in decomposition can improve the success rate for high-SNR spectra, it also decreases the success rate for low-SNR spectra. Moreover, the limited representativeness of the C1 bin quasar PCA templates, based on only 109 quasars, could be a significant factor contributing to the lower success rate.

The lower panel of Fig. \ref{fig:efficacy} displays the spectral SNR of reconstructed model for all successfully decomposed objects as a function of the SNR of its spectral data. Ideally, if the model is not overfitted to spectral noise, we would expect a uniform template SNR in the reconstructed model, regardless of the observed spectral SNR. The figure clearly shows that the template SNR from all three methods remains stable around 160 at the high SNR end, indicating that the PCA templates used in this work are effective in recovering spectra with SNRs greater than 10. At the low SNR end, the template SNR from linear decomposition methods decreases sharply, confirming the overfitting issue previously discussed in \S\ref{subsec:limitation}. In contrast, the template SNR based on the prior-informed method remains stable at a high level, indicating that the overfitting issue is significantly reduced.

\subsubsection{Host Fraction Accuracy} \label{subsubsec:hostfrac}

\begin{figure}[tb!]
   \includegraphics[width=0.48\textwidth]{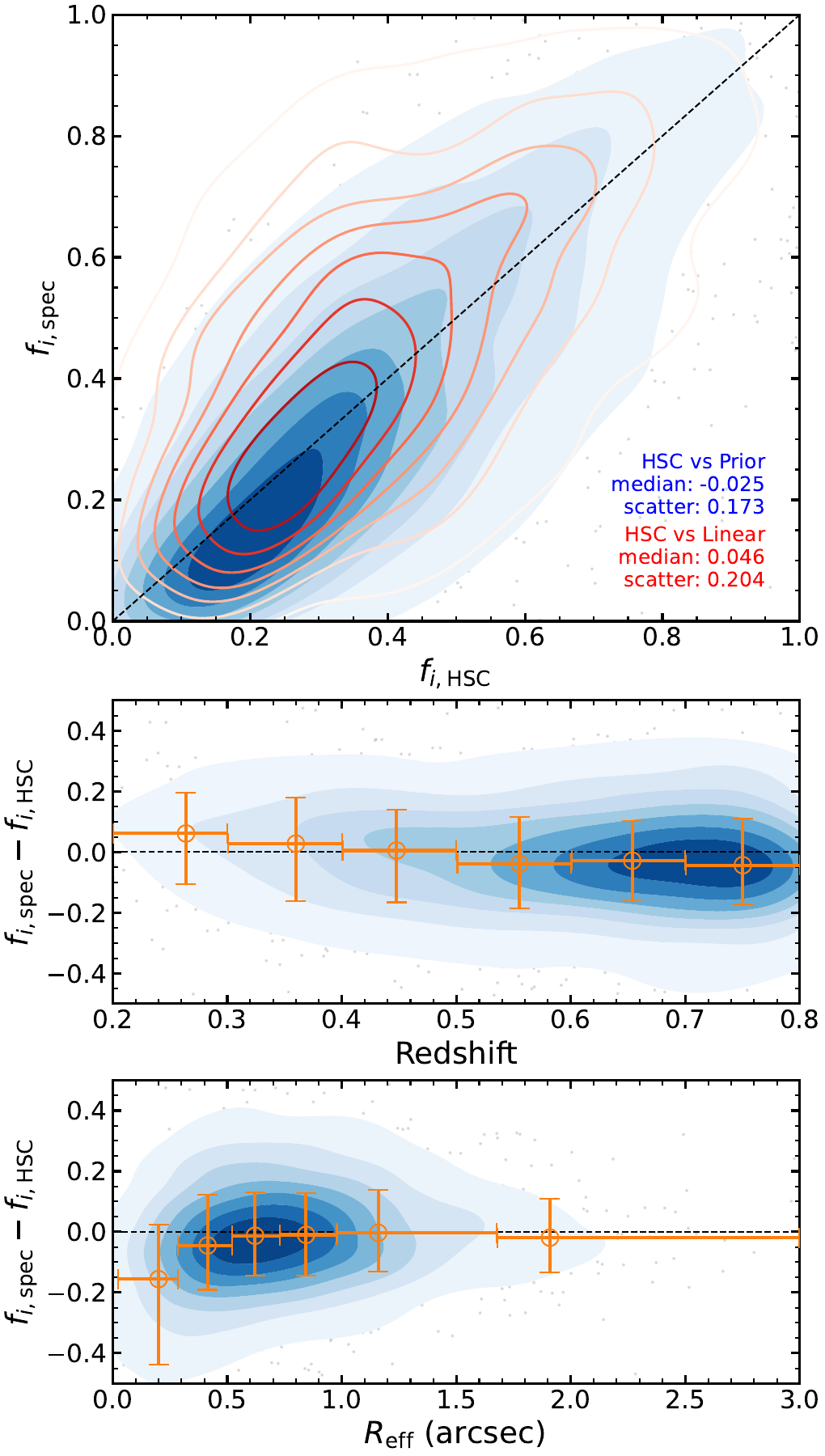}
   \caption{The comparison of $i$-band host fractions between the Hyper Suprime-Cam Subaru Strategic Program (HSC-SSP) image decomposition \citep{Li2021} and this study in the observed frame. In the upper panel, we directly compare the results of 2651 quasars in our sample having HSC images. As a comparison, linear decomposition method is also plotted as red contours. The host fraction difference dependence on redshift and effective radius of host galaxy are shown in the middle and lower panels. We show the binned median with orange circles. Their x-axis error indicate the bin size and their y-axis error indicate the 16\% to 84\% range.
   \label{fig:HSC_fh}}
\end{figure}

To verify the precision of our host decomposition, we compare our results with those from high-quality image decomposition. \citet{Li2021} have compiled an AGN-host decomposition catalog from the Hyper Suprime-Cam Subaru Strategic Program (HSC-SSP). This catalog is a subsample of the SDSS DR14Q catalog \citep{Paris2018}. For comparison, we applied the HSC $i$-band filter transmission function to our decomposed host galaxy template and the original spectra to determine the spectral host fraction in the HSC $i$-band, $f_i$. For the HSC images, we convolved the HSC decomposition model with a 2-D Gaussian with an FWHM equal to the median plate seeing and then applied a circular aperture (2\arcsec\ diameter for BOSS spectra and 3\arcsec\ diameter for SDSS spectra) to each model to derive a fiber-corrected host fraction. This process yielded 2651 quasars with valid host fraction measurements in both catalogs, whereas only 1560 of them can be decomposed using the linear decomposition method due to its relatively low success rate. 

The upper panel of Fig. \ref{fig:HSC_fh} illustrates the comparison between the host decomposition results using the prior-informed method (blue contours) and the linear decomposition method (red contours). Employing the image decomposition results as a benchmark, the prior-informed method achieves a median offset of $-0.025$ and a scatter of 0.173, while the linear method records a median offset of 0.046 and a scatter of 0.204. This suggests that the prior-informed method has significantly improved the success rate of decomposition and reduced the scatter in host fraction measurements, albeit with a slightly median offset, which should originate from intrinsic uncertainties in different methods and datasets.

The two lower panels of Fig. \ref{fig:HSC_fh} demonstrate that the differences between spectroscopic and image decompositions generally show no dependence on the redshift and effective radius of the host galaxy. The drop-off in differences for host galaxies with an effective radius less than $0.5\arcsec$ suggests that image decomposition may compete with the AGN model in compact galaxies, leading to an overestimation of the host fraction. Such a discrepancy could be reason for the weak decreasing trend in these differences relative to redshift, which the galaxy size would be smaller at higher redshifts.

\subsubsection{Stellar Velocity Dispersion Accuracy} \label{subsec:sigma_assess}

\begin{figure}[tb!]
   \includegraphics[width=0.48\textwidth]{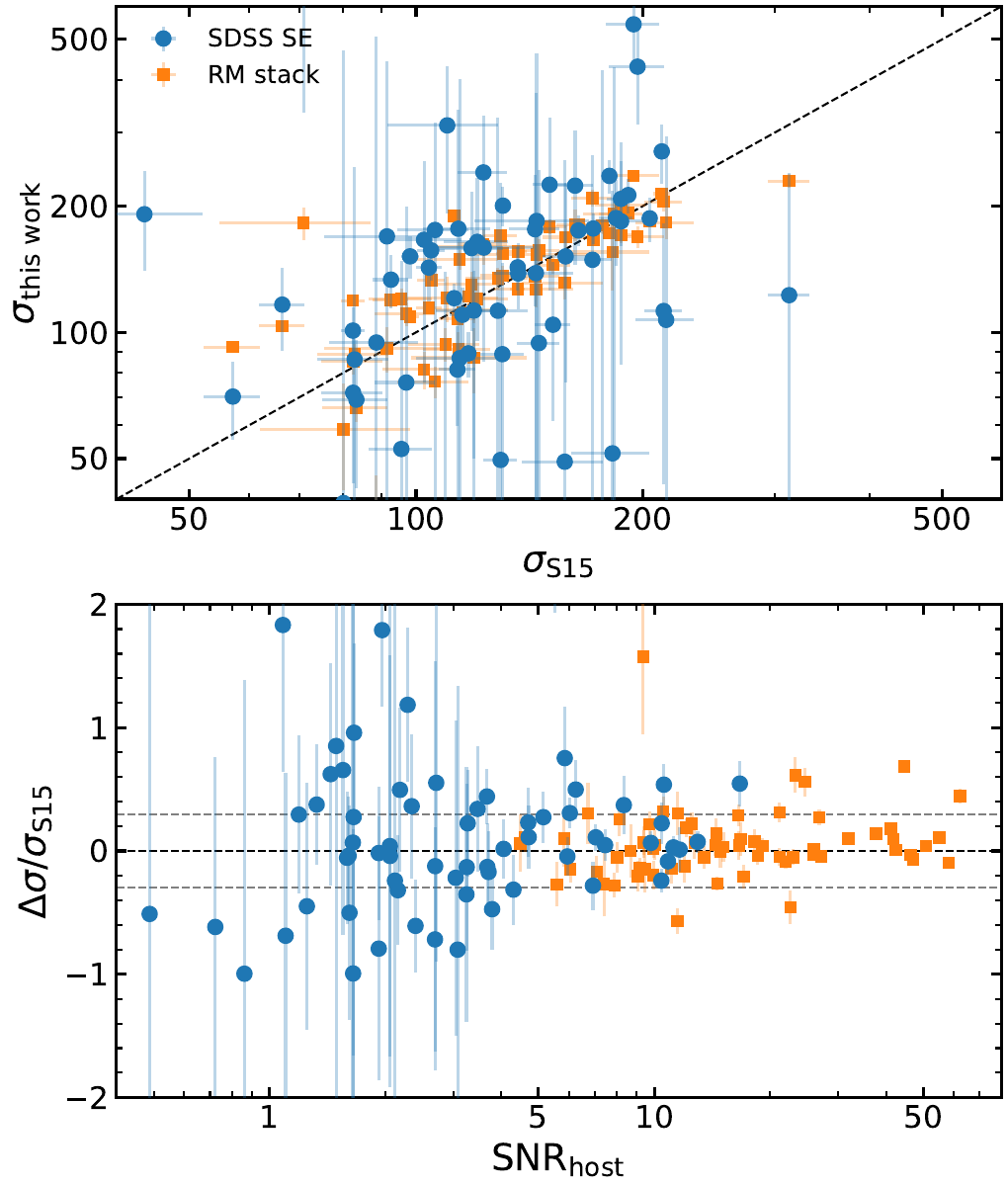}
   \caption{The comparison of stellar velocity measurements given by \citet{Shen2015a} and by this work. We apply our fitting routine to the best single-epoch spectra (SDSS SE) and the stacked high SNR spectra (RM stack) of these reverberation-mapping targets and give two sets of stellar velocity measurements. We present the direct comparison in the upper panel and the dependence of differences on the host galaxy spectra SNR in the lower panel. The gray dashed horizontal lines in the lower panel indicate the $\pm$ 30\% difference limits.
   \label{fig:sigma_cmp}}
\end{figure}

The stellar velocity dispersion $\sigma_\star$ is crucial for studying host galaxy properties. We evaluate the accuracy of our method in $\sigma_\star$ measurements by fitting the host component using the stellar kinematics fitting code {\tt pPXF} (see \S\ref{subsec:analysis} for details).

To assess the accuracy of our $\sigma_\star$ measurements, we compare our results with those of 66 high-quality RM stacked quasar hosts cross-matched from \citet{Shen2015a}. To evaluate the performance of our method across both low and high spectral SNRs, we apply our routine to both the stacked RM spectra and the corresponding best SE spectra (labeled as {\tt Science Primary} in SDSS archives) for each object. Fig. \ref{fig:sigma_cmp} presents our measured velocity dispersions, which are generally consistent with the stacked spectra (mostly within 30\%). However, the measurement is very sensitive to the spectral SNR. When switching to SE spectra with relatively lower SNRs, the inconsistency increases significantly, especially when spectral $\rm SNR_{host}<4$. Therefore, we suggest that a minimum criterion of ${\rm SNR_{host}}>4$ is required when using the $\sigma_\star$ measurements in our catalog.

\section{Application} \label{sec:apply}
The benchmark tests have demonstrated that our algorithm not only achieves a high success rate but also effectively mitigates overfitting issues. We thus applied our method to the entire low-redshift SDSS DR16 quasars. This section briefly outlines the data and spectral analysis methods employed. Subsequent discussions present the results of our measurements and explore the impact of host galaxy emission on existing AGN studies.

\subsection{Data and Sample Selection} \label{subsec:data}

The SDSS DR16Q catalog, compiled by \citet{Lyke2020a}, represents the final quasar sample from \SDSSIV and includes 750,414 quasars. The majority of the spectra were observed by BOSS spectrographs \citep{Smee2013}, which cover a wavelength range from 3600 to 10,400 \AA\ at a spectral resolution $R\sim$ 2000. The remaining spectra ($<$10\%) were obtained using \SDSSLEGACY\ spectrographs with a shorter wavelength coverage of 3800 to 9200 \AA. Given that the host galaxy's contribution is relatively low in distant, bright quasars, we limited our sample to a redshift of $z < 0.8$. This threshold ensures the inclusion of the \Hb\ emission line region and the 5100 \AA\ luminosity, which are critical for estimating $M_{\rm BH}$.  This criterion yields 76,565 quasars for further study.

\subsection{Spectral Analysis} \label{subsec:analysis}

Our fitting procedure is based on the open-source spectral fitting code {\tt PyQSOFit} \citep{2018ascl.soft09008G}. The prior-informed decomposition method has already been merged into the GitHub repository\footnote{https://github.com/legolason/PyQSOFit}. In the pipeline, the spectra are first corrected for Galactic extinction using a dust map by \citet{Schlegel1998} and a Milky Way extinction curve from \citet{Fitzpatrick1990}, and then transformed to the rest frame based on the redshift identified in DR16Q \citep{Lyke2020a}. Next, we utilize both the galaxy and quasar PCA spectra \citep{Yip2004, Yip2004a} to perform the prior-informed decomposition, separating the host galaxy and quasar components. For more details, also see \citet{Shen2011,Shen2019}.

\subsubsection{Measurements of the Host Galaxy properties} \label{subsubsec:host_measure}
The host galaxy spectra are derived by subtracting the reconstructed quasar eigenspectra from the original spectra (galaxy = data $-$ quasar templates). Due to the complexity of the broad component of the quasar spectra, the quasar PCA templates may not fully recover the broad line shape. To minimize the impact of residual broad \Ha\ and \Hb\ emission lines, we mask out the regions from 6400$\sim$6765 \AA\ and 4760$\sim$5035 \AA\ in the host galaxy spectra. For a conservative result, the pixel error from the original spectra is directly adopted as that of the host galaxy spectra in our subsequent measurement error estimations, which is apparently overestimated.

We then employ the stellar kinematics fitting code {\tt pPXF} to measure the stellar velocity dispersion ($\sigma_\star$) of the host galaxy. The Indo-U.S. Library of Coud\'{e} Feed Stellar Spectra \citep{Valdes2004a} was adopted as the stellar library, which covers the wavelength of 3460$\sim$9460 \AA\ at a spectral resolution of 1.35 \AA. We broaden the templates to the SDSS spectral resolution (2.76 \AA) before fitting. Only the first two moments (radial velocity and velocity dispersion) are fitted. Uncertainties are derived from the standard deviation of 100 mock spectra generated by adding Gaussian noise realization to the host galaxy spectra. 

Since our sample spans a large redshift range, to guarantee the consistency of our measurements, we restrict the fitting range to 4000 $-$ 5350 \AA, the same as used in \citet{Shen2015a}. A few significant stellar absorption features are within this range, including the G band (4304 \AA), the \MgIb\ triplet, and \ion{Fe}{1} blends (5270 \AA). The preferred infrared \CaII\ triplet around 8500 \AA\ falls outside the PCA templates used in this work. Additionally, the unreliable \CaHK\ line is excluded due to its strong dependence on spectral type, its location in a region with a steep local continuum, and significant intrinsic broadening, as noted in \citet{Greene2006a}.

We also measure the ${\rm D_n}4000$ index, defined as the flux ratio of two adjacent windows: 3850$\sim$3950 \AA\ and 4000$\sim$4100 \AA\ \citep{Balogh1999}. Its value increases as a function of stellar population age \citep[e.g.][]{Kauffmann2003, Zahid2015, Zahid2017}, thus serving as an age indicator of the host galaxy.

\subsubsection{Measurements of the Quasar Properties}

The quasar spectra are also derived by using the original spectra and subtracting the reconstructed galaxy eigenspectra (quasar = data $-$ galaxy templates). To avoid the blending of emission and absorption lines from hosts, we simply assume that the total emission line flux, e.g., \Hb\ and \OIII, is from quasars, ignoring the negligible contributions from the hosts.

The quasar spectral fitting procedure is similar to previous works \citep{Shen2019, Guo2019, Ren2022, Wu2022}. Here, we describe the general workflows and fitting setups. First, we fit the dereddened, deredshifted quasar spectra in several line-free regions using a power-law model along with a broadened empirical optical \FeII\ emission line template \citep{Boroson1992a}. After subtracting the continuum models, we fit the emission lines using Gaussian profiles on the emission-line-only spectrum. For the \Ha\ complex, we employ 3 Gaussian models to fit the broad components and 1 Gaussian for each narrow line including \Ha, the \NII\ doublet, and the \SII\ doublet. Similarly, for the \Hb\ complex, we use 3 Gaussian models for the broad \Hb\ component and 1 for narrow \Hb. A core component and a wind component are considered for each \OIII\ doublet. The relative line centers and line widths are fixed during the fitting for narrow lines and the wing components, respectively. 

\subsection{Results} \label{subsec:results}

We compiled our spectral measurements in Table \ref{tab:catalog}. Out of 76,565 low-redshift quasars, we successfully performed host decomposition for 71,760 quasars (93.7\%). In the following sections, we use the host fraction at 5100 \AA\ to represent the host contribution, unless specified otherwise. As a comparison, the decomposition success rate in \citet{Rakshit2020}, which uses the linear decomposition method, is only 23\% (12,807 out of 55,703). Next, we compare our refined measurements of AGN to those that do not properly separate the host-AGN components. 

\subsubsection{Host Galaxy Fraction $f_h$} \label{subsec:fh}

\begin{figure*}[tb!]
   \includegraphics[width=1\textwidth]{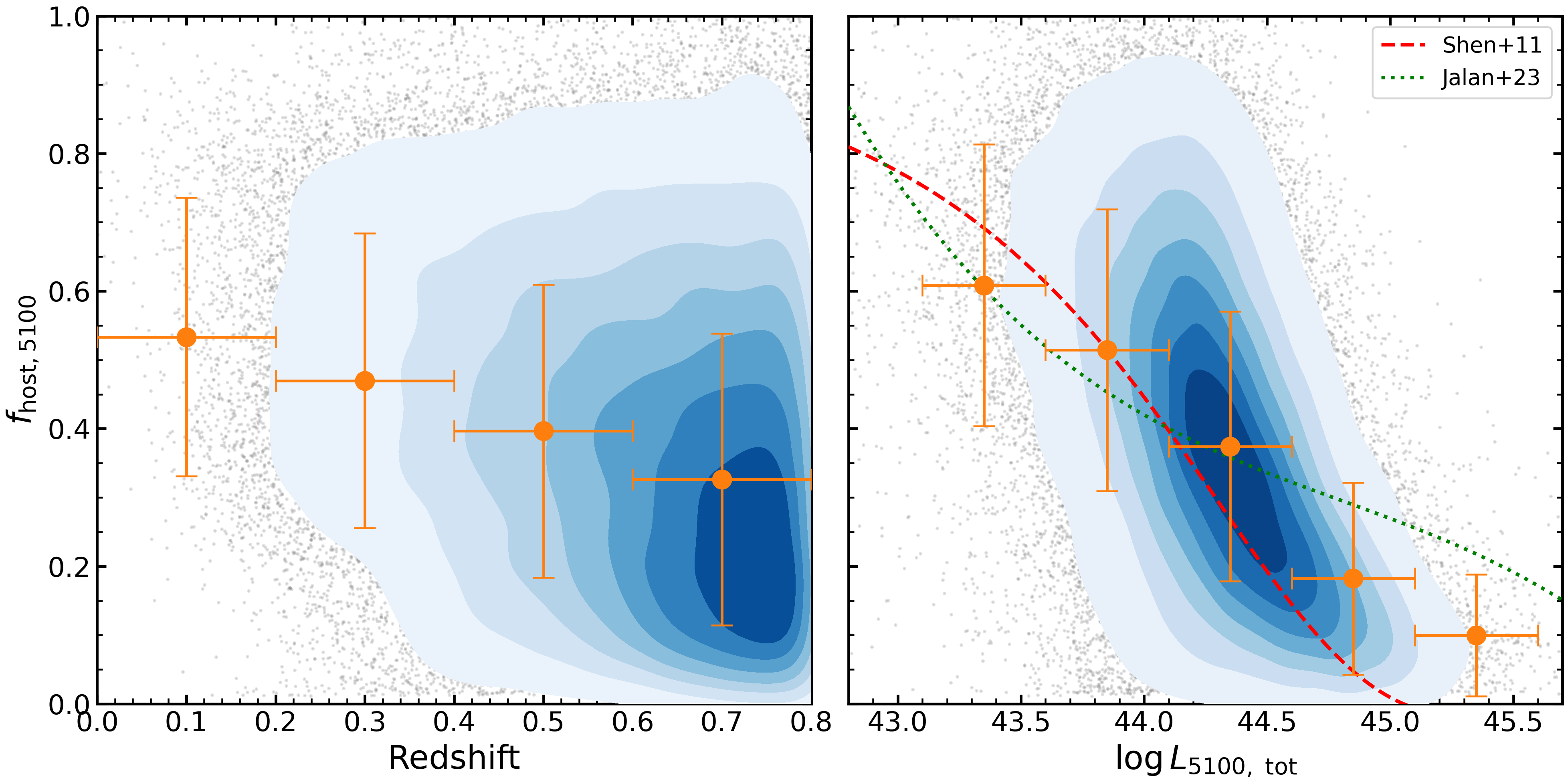}
   \caption{2D distributions (blue contours) of host fraction at 5100 \AA\ as a function of the redshift and total 5100 \AA\ luminosity. The orange dots show the $f_{\rm host,5100}$ distribution in different redshift and luminosity bins. The x-axis error bars indicate the bin range, and the y-axis error bars indicate the standard deviation of each bin. The red dashed and green dotted lines are previous results for comparison.
   \label{fig:fh_scatter}}
\end{figure*}

The host contribution in the continuum can be significant for low-redshift quasars in SDSS \citep[e.g.,][]{Jalan2023}. For all SDSS spectra with $z<0.8$, the median host contribution is $f_{h,5100}=35.7\%$, consistent with the results from SDSS-RM data, which report a median fraction of $38.8\%$ for a similar redshift range \citep{Shen2019}. 

Fig. \ref{fig:fh_scatter} displays the distributions of $f_{\rm host,5100}$ against redshift and total luminosity, $L_{\rm 5100,tot}$, along with the median for each x-axis bin. The host fraction gradually decreases from over 50\% at $z\sim0.1$ to approximately 30\% at $z\sim0.8$. This mild decline in host fraction with redshift could be due to a selection effect at higher redshifts, where only luminous quasars can be identified. Nevertheless, there is no effective redshift cut to avoid host contamination at optical wavelength.

The correlation between the host fraction and the total luminosity is more significant, with a decrease trend in median host fraction of $\sim65\%$ for the lowest luminosity bin ($L_{\rm 5100,tot}<43.6$) to $\sim10\%$ for the highest luminosity bin ($L_{\rm 5100,tot}>45.1$). This further confirms the selection effect that a less fractional host component decomposed in a more luminous target at a higher redshift. Two empirical relations suggested by \citet{Shen2011} and \citet{Jalan2023} are also plotted for comparison, which show similar trends to our results. The slight discrepancies among these studies are likely due to the using of different samples and decomposition methods. Generally, catalogs for low-redshift SDSS quasars should be used with caution if the host galaxy is not properly subtracted.

\subsubsection{Black Hole Masses} \label{subsec:LandM}

\begin{figure}[tb!]
   \includegraphics[width=0.48\textwidth]{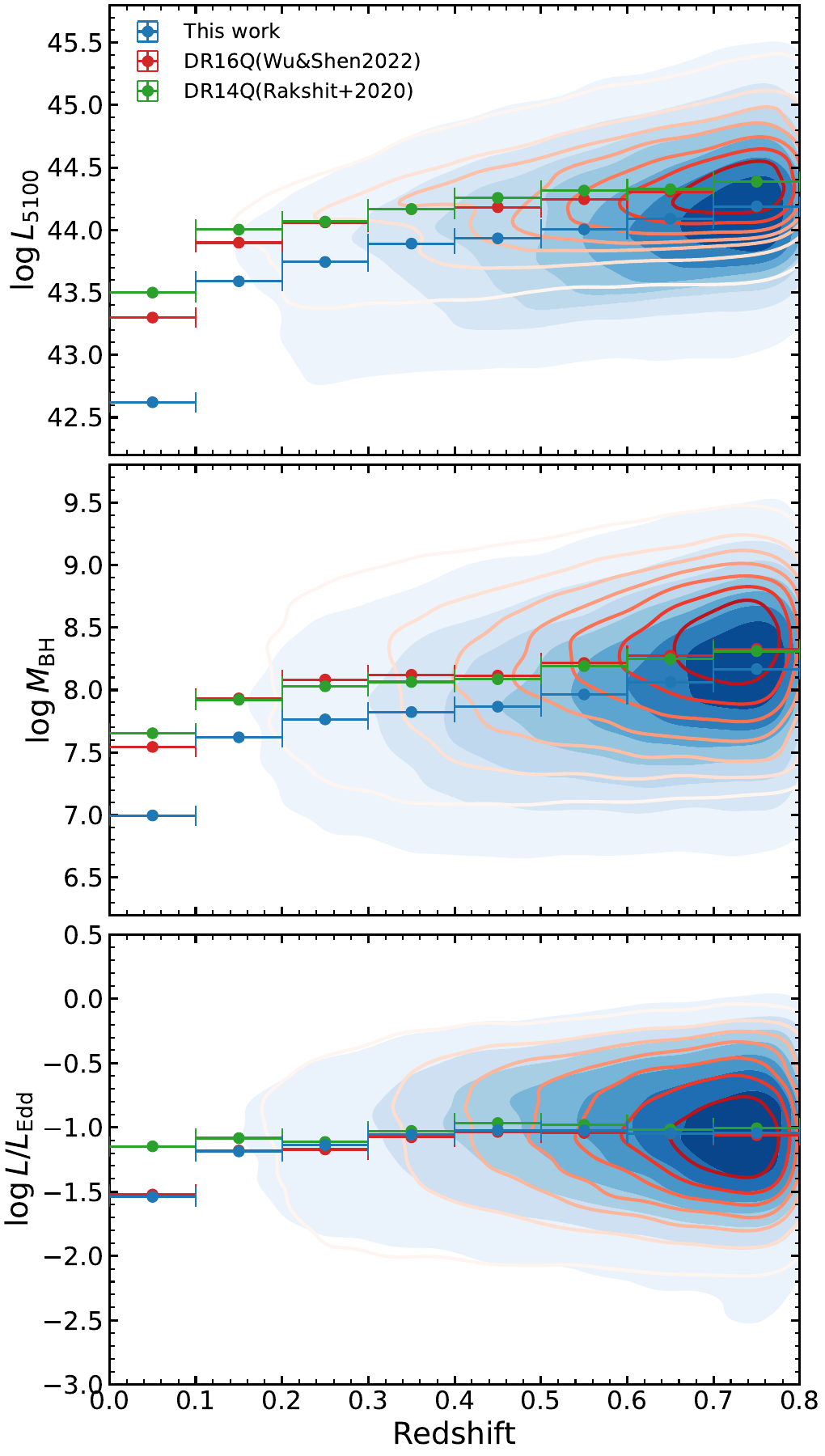}
   \caption{The distributions of measured $L_{5100}$, $M_{\rm BH}$, and Eddington ratio as a function of redshift are depicted for the DR16 quasar catalog \citep{Wu2022}, DR14 quasar catalog \citep{Rakshit2020}, and the current study. The filled blue contours represent our results, while the red contours correspond to DR16. Additionally, we bin the sample into intervals of 0.1 in redshift, displaying the corresponding median values for each bin. The difference in $L_{5100}$ comes from the host galaxy subtraction, while the difference in $M_{\rm BH}$ is a combined result of the host galaxy subtraction, the different mass recipe, and the improved FWHM measurements (See \S\ref{subsec:LandM}).
   \label{fig:DR16_cmp}}
\end{figure}

In this section, we provide the $M_{\rm BH}$ measurement based on our host-subtracted spectra, and quantitatively compare it with previous studies. Furthermore, we will discuss the potential influence of non-host-subtracted measurements on previously obtained results.

The black hole masses $M_{\rm BH}$ are usually calculated based on the 5100 \AA\ continuum luminosity $L_{5100}$ and \Hb\ broad emission line width ${\rm FWHM_{H\beta}}$, which is the so-called ``single-epoch virial BH mass" estimators:
\begin{equation}
\log{\left(\frac{M_{\rm BH}}{M_\odot}\right)} = a + b\log{\left(\frac{L_{5100}}{10^{44}~\ergs}\right)}+c\log{\left(\frac{\rm FWHM_{H\beta}}{\kms}\right)},
\end{equation}
where coefficients $a$, $b$, and $c$ are determined empirically using the RM AGN sample \citep[e.g.,][]{Vestergaard2006, Feng2014, Shen2019}, the fiducial mass recipe, commonly used by quasar studies \citep[e.g.,][]{Shen2011, Rakshit2020, Wu2022}, originates from \citet{Vestergaard2006} and specifies coefficients $a=0.91$, $b=0.50$, and $c=2.0$. However, it does not correct for host contributions. Therefore, in this work, we adopted the host-corrected recipe derived by \citet[][hereafter S23]{Shen2023a}, which modifies $a$ to 0.85 only. The statistical uncertainties of $M_{\rm BH}$ are given by the propagation of uncertainties in $L_{5100}$ and ${\rm FWHM_{H\beta}}$. For individual quasars, an absolute systematic uncertainty of approximately 0.35 dex in SE masses should be considered \citep{Shen2023a}. The bolometric luminosity is estimated using $L_{\rm bol} = 9.26 \times L_{5100}$, where 9.26 is the bolometric correction obtained from \citet{Richards2006}. The Eddington ratio $\lambda_{\rm Edd} = L_{\rm bol}/L_{\rm Edd}$, where $L_{\rm Edd}$ is the Eddington luminosity.

We then compare our revised $L_{5100}$ and $M_{\rm BH}$ measurements with those from \citet{Wu2022}, which did not account for host contributions. Fig. \ref{fig:DR16_cmp} displays the distributions of these two samples as a function of redshift, overlapped the median value for each redshift bin. Considering host emissions, our results show a median $L_{5100}$ that is 0.215 dex lower and a median $M_{\rm BH}$ that is 0.219 dex lower compared to the values reported in \citet{Wu2022}. Despite these differences, the Eddington ratios between the two samples remain consistent, likely due to coincidence. Fig. \ref{fig:DR16_cmp} also includes the distribution from the SDSS DR14Q property catalog \citep{Rakshit2020} for comparison, which adopted the linear decomposition method with global quasar PCA templates. The discrepancies observed may be attributed to differences in the samples and the low success rate of decomposition in DR14Q.

Several factors contribute to the systematically lower $M_{\rm BH}$, measured as 0.219 dex lower in our study. Firstly, the constant $a$ in the S23 recipe is 0.06 dex smaller than that in VP06. Additionally, the subtraction of the host contribution, 38.8\% on average (see \S\ref{subsec:fh}), accounts for another 0.11 dex decrease in $M_{\rm BH}$. The remaining 0.05 dex discrepancy is likely attributable to degeneracy in the spectral decomposition process in different works.

Accurate determination of $M_{\rm BH}$ is crucial for deepening our understanding of AGN dynamics and their coevolution with the surrounding environment. Given the systematic overestimation of $M_{\rm BH}$ in low-redshift quasars, we suggest that previous studies relying on the absolute value of $M_{\rm BH}$ could be biased. For instance, the ratio of $M_{\rm BH}$ to galaxy stellar mass based on SE spectra without host correction \citep[e.g.,][]{Ding2020, Li2021a}, might be overestimated compared to local results, where $M_{\rm BH}$ is dynamically measured and relatively free from host contamination. A similar issue exists in estimating the black hole mass function (BHMF): Previous studies often set a redshift limit of $z>0.3$ to minimize host contamination on $M_{\rm BH}$ measurements \citep{Shen2012, Kelly2013}, which, according to our results, is not sufficient. Considerable bias in $M_{\rm BH}$ estimation persists even up to $z\sim0.8$, raising concerns for BHMF studies using SE mass estimations.

\subsubsection{$M_{\rm BH}$-$\sigma_\star$ Relation}

\begin{figure}[tb!]
   \includegraphics[width=0.44\textwidth]{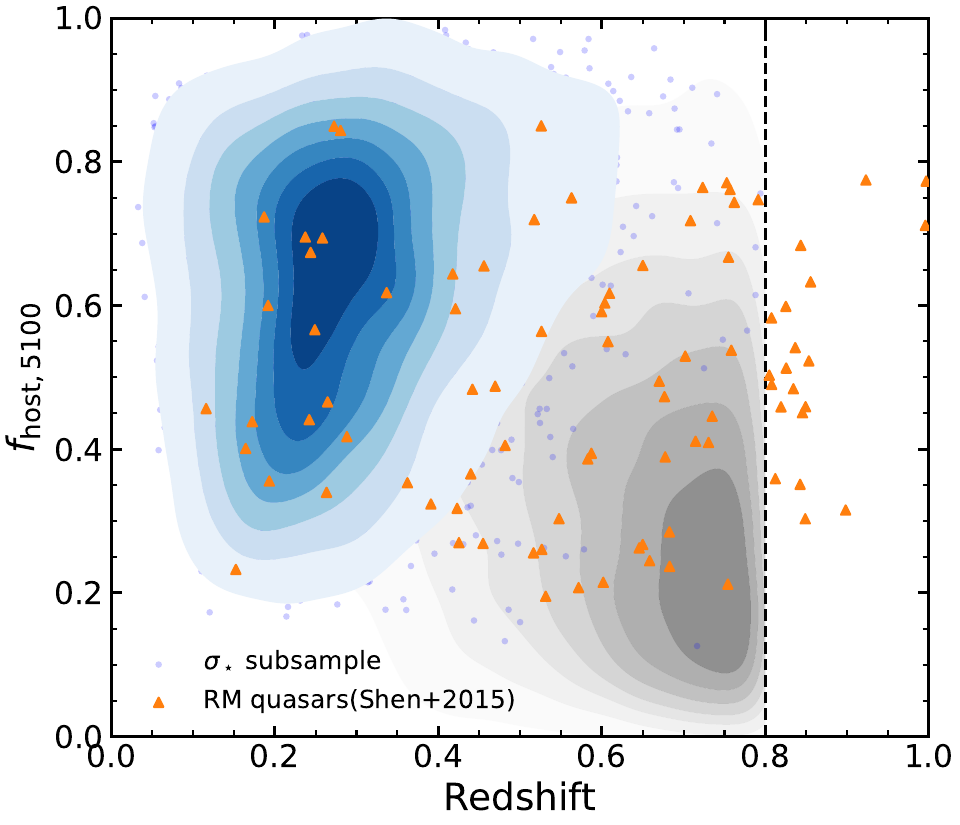}
   \includegraphics[width=0.44\textwidth]{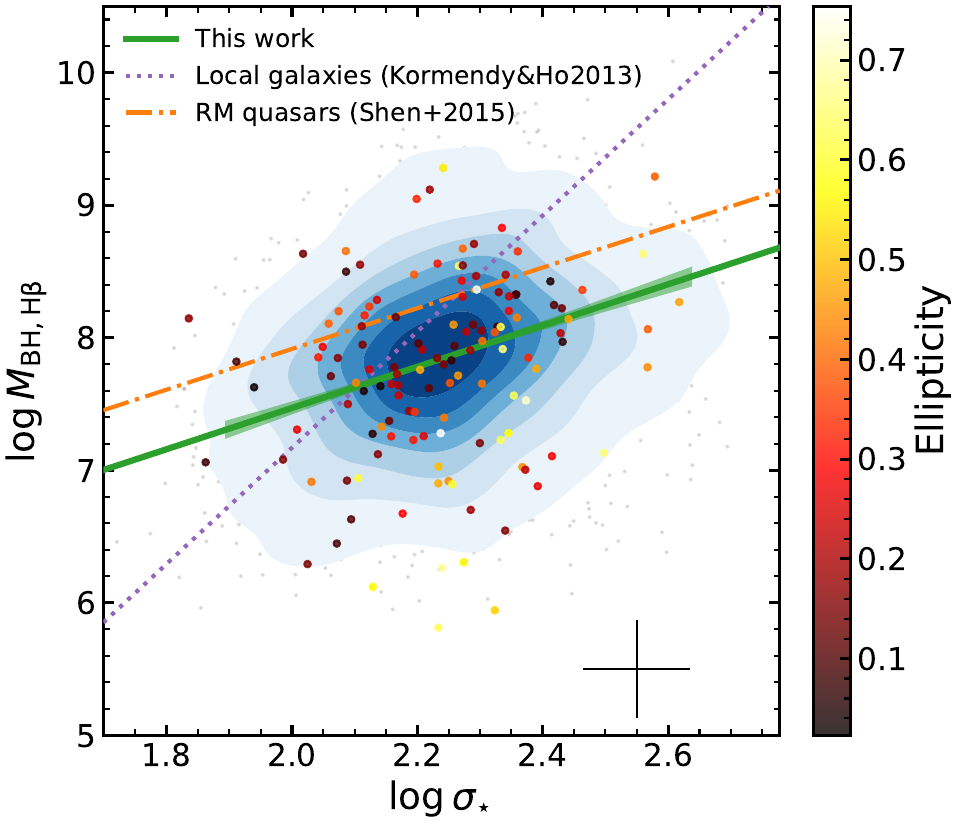}
   \caption{Upper panel: sample distribution in the $f_{\rm host,5100}$-redshift plane. Blue contours and dots represent the quasar subsample that passed the $\sigma_\star$ quality cuts. The distribution of our main sample is shown in gray contours with a vertical dashed line indicating the redshift cut at 0.8. The quasars used in \citet{Shen2015a} are shown as orange triangles for comparison. Lower panel: the blue contours show the $M_{\rm BH}-\sigma_\star$ relation for the $\sigma_\star$ quality cut subsample with its average uncertainty at the lower right corner. The green line and the dashed region indicate the best-fit relation and the 95\% confidence band derived by {\tt linmix}. We add the relation derived by \citet{Kormendy2013, Shen2015a} for reference. We also cross-matched 151 quasars with HSC image in \citet{Li2021a} and color coded with the ellipticity of the host galaxy to illustrate the galaxy inclination effect on $\sigma_\star$ measurements.
   \label{fig:Msigma}}
\end{figure}

Given the high success rate of our host decomposition method, we are able to investigate the $M_{\rm BH}$-$\sigma_\star$ relation across a larger sample. To ensure the measurement reliability of the stellar velocity dispersion $\sigma_\star$, we use the same criteria as \citet{Shen2015a}: 1) the host fraction at 4200 \AA\ is required to be larger than 0.05; 2) The spectral SNR of host galaxy should be larger than 4; 3) $\sigma_\star$ is above $3\sigma$ confidence level (see \S\ref{subsec:sigma_assess} for the discussion of the reliability of our $\sigma_\star$ measurements).

In total, 4137 quasars passed these quality cuts. The distribution of this subsample in the $f_{\rm host,5100}$-redshift plane is shown in the upper panel of Fig. \ref{fig:Msigma}. Different from the main sample, this selection criteria favor quasars with relatively higher host fraction and lower redshift. The median redshift and the median host fraction of this subsample are 0.27 and 63\%, respectively. Although we discard most of the high-redshift quasars, this subsample still has a reasonable completeness at low redshift (e.g., 45\% for $z<0.3$).

Start from $\sigma_\star$ quality cut subsample, we further studied the relation between their $\sigma_\star$ and $M_{\rm BH}$ based on broad \Hb\ line. We use {\tt linmix}, a Bayesian linear regression code based on the method developed by \citet{Kelly2007}, to perform the linear fit:
\begin{equation}
\log{\left(\frac{M_{\rm BH,H\beta}}{M_\odot}\right)}=\alpha+\beta\log{\left(\frac{\sigma_\star}{200~\kms}\right)}.
\end{equation}
To account for the scatter due to the systematic errors in SE mass estimates, we additionally add 0.35 dex to the measured statistical errors \citep{Shen2023a} during the fitting. The lower panel of Fig. \ref{fig:Msigma} presents the best-fit linear relation, with an intercept of $\alpha=7.940 \pm 0.010$ and a slope of $\beta=1.231 \pm 0.069$ (green line). The intrinsic scatter of $\log{M_{\rm BH}}$ around the best-fit model is 0.129. 

Our result holds a similar slope to that from stacked RM quasar sample \citep[orange dash-dotted line,][]{Shen2015a}, yet with $\sim0.5$ dex lower in $M_{\rm BH}$. The difference in intercept is primarily due to the sample differences in redshift and host fraction, shown in the upper panel of Fig. \ref{fig:Msigma}. Since the quasars spectra in \citet{Shen2015a} are stacked from about 38 epochs with the accumulated exposure time tens of times higher than that of our SE spectra, their high SNR allows a higher completeness at higher redshift and lower host fraction under the same selection criteria. On the contrary, our sample would favor quasars with relatively lower AGN luminosity and higher galaxy luminosity, leading to a systematic lower-right shift in Fig.~\ref{fig:Msigma}. Besides, the new recipe for SE $M_{\rm BH}$ estimation and the reduction of host contamination would also contribute to $\sim0.2$ dex discrepancy, as clarified in \S\ref{subsec:LandM}.

Our $M_{\rm BH}$-$\sigma_\star$ relation is significantly shallower than the local relation for inactive galaxies \citep[purple dotted line,][]{Kormendy2013}. The divergence from the local relation is more complex. One frequently cited factor is the luminosity constraints, under which more luminous quasars with brighter broad lines are preferentially identified. Assuming a positive correlation between luminosity and $M_{\rm BH}$, this selection bias leads to a more massive $M_{\rm BH}$-biased sample with respect to $\sigma_\star$ \citep{Lauer2007a}. In a flux-limited sample, such an offset becomes more pronounced at lower $\sigma_\star$ values, resulting in a shallower slope in the $M_{\rm BH}-\sigma_\star$ relation \citep{Shen2015a}. Note that the redshift of our sample is generally lower than 0.4, so the bias described by \citet{Lauer2007a} would be less severe than that in the sample of \citet{Shen2015a}.

The decomposition procedure would introduce additional uncertainties to the $\sigma_\star$ measurements beyond the statistical error. Passing underestimated error could lead to a shallower fitted slope

The accuracy of $\sigma_\star$ measurements will also impact the fitted slope. Underestimating the $\sigma_\star$ error due to the decomposition procedure, beyond mere statistical errors, and passing this underestimated error to the fitting process could lead to a shallower fitted slope. Besides, the $\sigma_\star$ estimation of a subset of our sample could be overestimated, which would extend the true $\sigma_\star$ distribution to the right in Fig. \ref{fig:Msigma}, and therefore shallow the fitted slope. The measured integrated velocity dispersion from the spectrum is contributed to by both the inner bulge stellar velocity dispersion and the rotational velocity from the outer disk. As will be discussed in \S\ref{subsubsec:host_prop} that the AGN hosts tend to be star-forming galaxies, which are likely to be disk dominated. Therefore, the measured dispersion will represent the intrinsic $\sigma_\star$ only for face-on galaxies, while being boosted by the rotational velocity for edge-on galaxies, leading to an overestimation of $\sigma_\star$ \citep{Bezanson2015}. Fig. \ref{fig:Msigma} highlights 143 quasars with morphology measurements in \citet{Li2021a}, color coded by the ellipticity ($\epsilon = 1 - b/a$) of the host galaxy, which is correlated with the inclination angle. Galaxies with higher ellipticity tend to reside on the right side of the local relation, indicating a possible overestimation of $\sigma_\star$.

The potential evolution of the $M_{\rm BH}-\sigma_\star$ relation may also play a role in this divergence. \citet{Silverman2022} suggest an intrinsic shallower $M_{\rm BH}-\sigma_\star$ slope at higher redshift based on the SDSS quasar at $0.2<z<0.8$, with selection bias corrected. Since the quasars in our sample have a median $z\sim0.3$, the change of intrinsic relation can be a possible reason.

\subsubsection{Stellar Age of Quasar Hosts} \label{subsubsec:host_prop}

\begin{figure}[tb!]
   \includegraphics[width=0.48\textwidth]{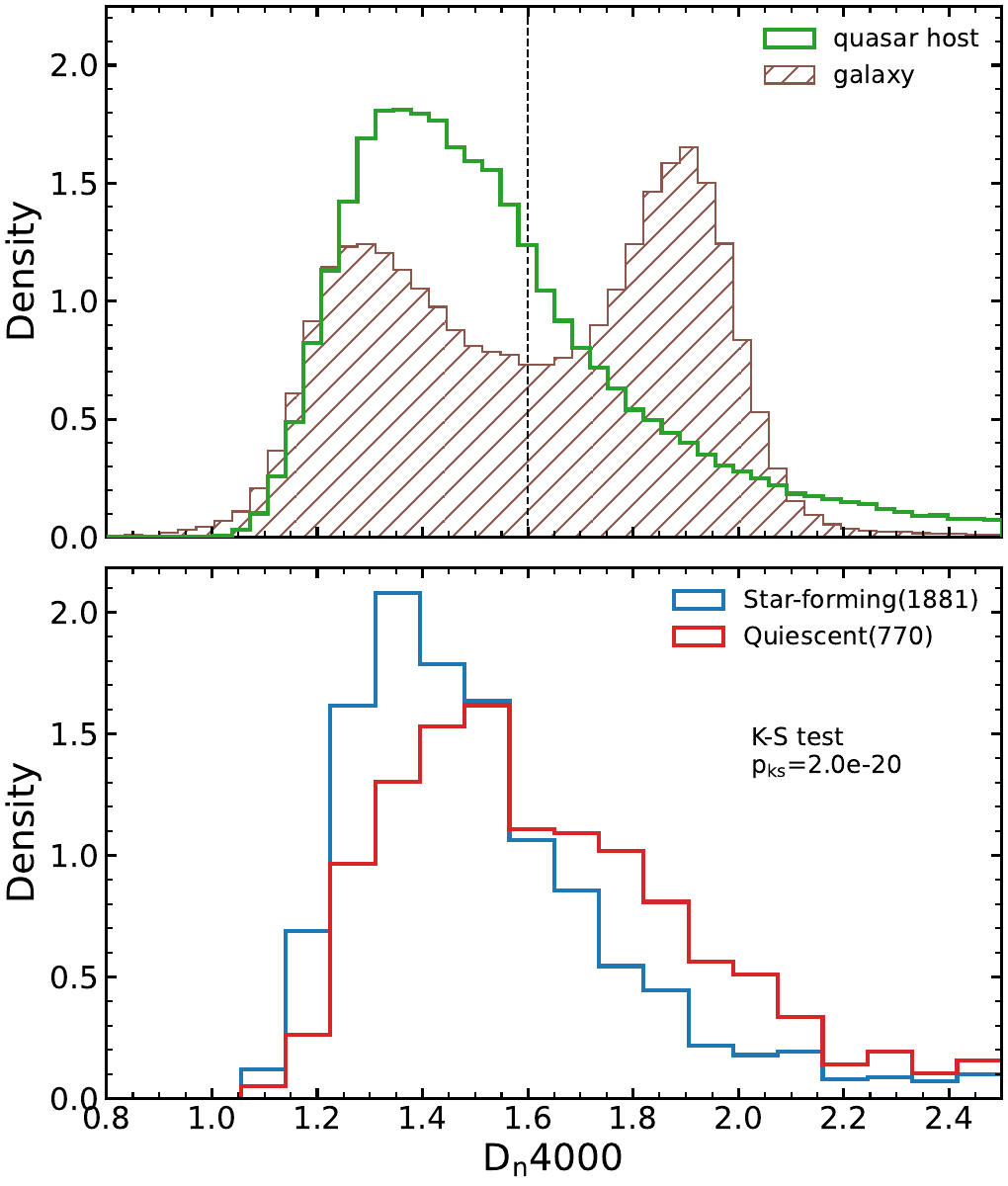}
   \caption{The upper panel shows the distribution of ${\rm D_n}$4000 index for our decomposed quasar hosts. The distribution of the ${\rm D_n}4000$ index for the galaxy sample used to build the galaxy PCA template is also plotted (hatched histogram) for comparison. Quasar hosts generally reside on the left side of the distribution, consistent with the younger age of the star-forming galaxy population. Quasar hosts with HSC-based $\rm{UVM_\star}$ diagram classification \citep{Li2021a} are displayed in the lower panel, separated by image-based galaxy type. The number of quasars in each bin is listed in the legend.
   \label{fig:Dn4000}}
\end{figure}

Previous studies have demonstrated that quasars are predominantly hosted by young galaxies, indicating that the BH accretion rate tracks the star formation rate over cosmic history \citep[e.g.,][]{Kormendy2013, Matsuoka2013, Trump2013, Xie2021}. The ${\rm D_n}$4000 index, a commonly used age indicator, measures the break strength at 4000 \AA\ and reflects the age of host galaxies. This index is typically smaller than 1.6 for young stellar populations and higher than 1.6 for older ones \citep{Kauffmann2003}. The upper panel of Fig. \ref{fig:Dn4000} clearly illustrates a young distribution among our quasar hosts (green histogram), generally consistent with the host ages of X-ray selected Type 2 AGNs that has little contamination from AGN continuum \citep{Silverman2009, Ni2023}. Our results not only confirm the existing findings that cold gas serves as a common fuel for the SMBH in quasars, but also validate our host decomposition method as reliable for obtaining host properties.

We further checked our measured ${\rm D_n}$4000 index of star-forming and quiescent galaxies, identified by HSC image-based $\rm{UVM_\star}$ diagram \citep{Li2021a}, in our sample. The lower panel of Fig. \ref{fig:Dn4000} shows a significant difference between two groups, with a K-S test p-value of $2.0\times10^{-20}$. Despite a considerable overlap, this classification of the star-forming and quiescent galaxy populations can be verified by the ${\rm D_n}$4000 index from our method, further demonstrating the credibility of our host galaxy decomposition.

\section{Summary} \label{sec:summary}

We introduced a prior-informed AGN-host spectra decomposition technique which is now implemented in {\tt PyQSOFit}, which significantly increased the decomposition success rate and provided accurate AGN-host flux ratios, host-free $L_{\rm bol}$ and $M_{\rm BH}$ estimations and host galaxy properties ($\sigma_\star$ and ${\rm D_n}$4000). We applied this method to SDSS DR16 quasars containing 76,565 quasars at $z<0.8$, and, for the first time, compiled a survey-level, host-decomposed catalog.

The main novelty of our host decomposition is to use the distribution of the PCA eigencoefficient as prior information to decrease the degeneracy of PCA templates between quasar and galaxy datasets, and efficiently reduced the overfitting issue. Our decomposition success rate on SDSS DR16 quasar catalog reaches 93.7\%, substantially higher than previous works (Fig. \ref{fig:efficacy}). 

We illustrate that the average host galaxy contribution in these low-$z$ quasar spectra is 38.8\% at 5100 \AA. If the host galaxy is not taken into account, this will result in an overestimation of 0.215 dex in $L_{5100}$ and 0.219 dex in$ M_{\rm BH}$ (Fig. \ref{fig:DR16_cmp}), which could lead to biased conclusions for previous related works, such as the $M_{\rm BH}-M_{\star}$ relation and black hole mass function.

We suggest that the $\sigma_\star$ measurements are reliable with a scatter of 0.15 dex when host spectral SNR are larger than 4 (Fig. \ref{fig:sigma_cmp}). The $M_{\rm BH}-\sigma_\star$ relation in our sample is significantly shallower than the local relation \citep{Kormendy2013}, however, consistent to that from high-redshift quasar sample \citep{Shen2015a}. We argue that both sample selection bias and measurement scatter would affect the fitted slope. 

We note that, limited by the sample size and the spectral resolution of current PCA templates, our code is not optimized for quasars with intrinsically low luminosity or high spectral resolution. To adapt this method for compatibility with next generation of spectroscopic surveys such as DESI \citep{DESICollaboration2016,DESICollaboration2016a}, 4MOST \citep{Jong2014}, and PFS \citep{Takada2014}, a new set of PCA or independent component analysis \citep[][]{Lu2006} templates based on a more complete sample with higher spectral resolution is required for future work.

All products from this work, including the catalog, fitted parameters and quality assurance figure can be accessed from \url{https://zenodo.org/doi/10.5281/zenodo.11376764}.


\startlongtable
\centerwidetable
\begin{deluxetable*}{rlccp{0.45\linewidth}}
\tablecaption{FITS catalog format}\label{tab:catalog}
\tablehead{
    \colhead{Number} & \colhead{Column Name} & \colhead{Format} & \colhead{Unit} & \colhead{Description}
}
\decimalcolnumbers
\startdata
        1 & SDSS\_NAME             & string      &                    & Unique identifier from the SDSS DR16 quasar catalog      \\
        2 & RA                    & float32     & deg                & Right ascension (J2000)                                  \\
        3 & DEC                   & float32     & deg                & Declination (J2000)                                      \\
        4 & Z                     & float32     &                    & Redshift                                                 \\
        5 & PLATE                 & int32       &                    & SDSS plate number                                        \\
        6 & MJD                   & int32       &                    & MJD when spectrum was observed                           \\
        7 & FIBER                 & int32       &                    & SDSS fiber ID                                            \\
        8 & SN\_CONTI              & float32     &                    & Mean signal-to-noise ratio estimated at 5100 $\rm{\AA}$    \\
        9 & LOGL5100\_TOT          & float32     & \ergs            & Logarithmic total luminosity at rest-frame 5100 $\rm{\AA}$ \\
       10 & LOGL5100\_TOT\_ERR      & float32     & \ergs            & Error in LOGL5100\_TOT                                    \\
       11 & RCHI2\_DECOMP          & float32     &                    & Reduced chi-square of the PCA decomposition              \\
       12 & HOST\_FR\_5100               & float32     &                    & Fraction of host galaxy at rest-frame 5100 $\rm{\AA}$      \\
       13 & HOST\_FR\_4200               & float32     &                    & Fraction of host galaxy at rest-frame 4200 $\rm{\AA}$      \\
       14 & HOST\_FR\_REST\_G             & float32     &                    & Fraction of host galaxy at rest-frame SDSS g-band        \\
       15 & HOST\_FR\_REST\_R             & float32     &                    & Fraction of host galaxy at rest-frame SDSS r-band        \\
       16 & HOST\_FR\_OBS\_G              & float32     &                    & Fraction of host galaxy at observer-frame SDSS g-band    \\
       17 & HOST\_FR\_OBS\_R              & float32     &                    & Fraction of host galaxy at observer-frame SDSS r-band    \\
       18 & HOST\_FR\_OBS\_I              & float32     &                    & Fraction of host galaxy at observer-frame SDSS i-band    \\
       19 & HOST\_REST\_GMAG         & float32     & mag                & SDSS g-band host galaxy magnitude at rest-frame          \\
       20 & HOST\_REST\_RMAG         & float32     & mag                & SDSS r-band host galaxy magnitude at rest-frame          \\
       21 & HOST\_OBS\_GMAG         & float32     & mag                & SDSS g-band host galaxy magnitude at observer-frame      \\
       22 & HOST\_OBS\_RMAG         & float32     & mag                & SDSS r-band host galaxy magnitude at observer-frame      \\
       23 & HOST\_OBS\_IMAG         & float32     & mag                & SDSS i-band host galaxy magnitude at observer-frame      \\
       24 & SN\_HOST               & float32     &                    & Signal-to-noise ratio of the host galaxy                 \\
       25 & DN4000                & float32     &                    & ${\rm D_n}$4000 index of the host galaxy                          \\
       26 & VDISP                 & float32     &                    & Velocity dispersion of the host galaxy measured by pPXF  \\
       27 & VDISP\_ERR             & float32     &                    & Error in VDISP                                           \\
       28 & VOFF                  & float32     &                    & Velocity offset of the host galaxy measured by pPXF      \\
       29 & VOFF\_ERR              & float32     &                    & Error in VOFF                                            \\
       30 & RCHI2\_PPXF            & float32     &                    & Reduced chi-square of the pPXF fitting                   \\
       31 & PL\_SLOPE              & float32     &                    & Slope of AGN power law                                   \\
       32 & PL\_SLOPE\_ERR          & float32     &                    & Error in PL\_SLOPE                                        \\
       33 & LOGL5100              & float32     & \ergs            & AGN continuum luminosity at rest-frame 5100 $\rm{\AA}$     \\
       34 & LOGL5100\_ERR          & float32     & \ergs            & Error in LOGL5100                                        \\
       35 & LOGL4200              & float32     & \ergs            & AGN continuum luminosity at rest-frame 4200 $\rm{\AA}$     \\
       36 & LOGL4200\_ERR          & float32     & \ergs            & Error in LOGL4200                                        \\
       37 & LOGL3000              & float32     & \ergs            & AGN continuum luminosity at rest-frame 3000 $\rm{\AA}$     \\
       38 & LOGL3000\_ERR          & float32     & \ergs            & Error in LOGL3000                                        \\
       39 & FE\_4435\_4685\_FLUX     & float32     & $10^{-17}$ \ergscm & Flux of Fe II complex within the 4435-4685 $\rm{\AA}$      \\
       40 & FE\_4435\_4685\_FLUX\_ERR & float32     & $10^{-17}$ \ergscm & Error in FE\_4435\_4685\_FLUX                               \\
       41 & LOGLBOL               & float32     & \ergs            & Bolometric luminosity estimated based on LOGL5100        \\
       42 & LOGLBOL\_ERR           & float32     & \ergs            & Error in LOGLBOL\_5100                                    \\
       43 & LOGMBH                & float32     & $M_\odot$              & Black hole mass estimated based on broad \Hb\ line         \\
       44 & LOGMBH\_ERR            & float32     & $M_\odot$              & Error in LOGMBH\_HB                                       \\
       45 & LOGLEDD\_RATIO               & float32     &                    & Logarithmic Eddington ratio                              \\
       46 & HA\_BR\_SNR             & float32     &                    & Signal-to-noise ratio of broad \Ha\ component              \\
       47 & HA\_BR\_PEAK            & float32     & $\rm{\AA}$           & Peak wavelength of broad \Ha\ component                    \\
       48 & HA\_BR\_PEAK\_ERR        & float32     & $\rm{\AA}$           & Error in HA\_BR\_PEAK                                      \\
       49 & HA\_BR\_FLUX            & float32     & $10^{-17}$ \ergscm & Flux of broad \Ha\ component                               \\
       50 & HA\_BR\_FLUX\_ERR        & float32     & $10^{-17}$ \ergscm & Error in HA\_BR\_FLUX                                      \\
       51 & HA\_BR\_EW              & float32     & $\rm{\AA}$           & Rest-frame EW of broad \Ha\ component                      \\
       52 & HA\_BR\_EW\_ERR          & float32     & $\rm{\AA}$           & Error in HA\_BR\_EW                                        \\
       53 & HA\_BR\_SIGMA           & float32     & \kms             & Line dispersion of broad \Ha\ component                    \\
       54 & HA\_BR\_SIGMA\_ERR       & float32     & \kms             & Error in HA\_BR\_SIGMA                                     \\
       55 & HA\_BR\_FWHM            & float32     & \kms             & FWHM of broad \Ha\ component                               \\
       56 & HA\_BR\_FWHM\_ERR        & float32     & \kms             & Error in HA\_BR\_FWHM                                      \\
       57 & HB\_BR\_SNR             & float32     &                    & Signal-to-noise ratio of broad \Hb\ component              \\
       58 & HB\_BR\_PEAK            & float32     & $\rm{\AA}$           & Peak wavelength of broad \Hb\ component                    \\
       59 & HB\_BR\_PEAK\_ERR        & float32     & $\rm{\AA}$           & Error in HB\_BR\_PEAK                                      \\
       60 & HB\_BR\_FLUX            & float32     & $10^{-17}$ \ergscm & Flux of broad \Hb\ component                               \\
       61 & HB\_BR\_FLUX\_ERR        & float32     & $10^{-17}$ \ergscm & Error in HB\_BR\_FLUX                                      \\
       62 & HB\_BR\_EW              & float32     & $\rm{\AA}$           & Rest-frame EW of broad \Hb\ component                      \\
       63 & HB\_BR\_EW\_ERR          & float32     & $\rm{\AA}$           & Error in HB\_BR\_EW                                        \\
       64 & HB\_BR\_SIGMA           & float32     & \kms             & Line dispersion of broad \Hb\ component                    \\
       65 & HB\_BR\_SIGMA\_ERR       & float32     & \kms             & Error in HB\_BR\_SIGMA                                     \\
       66 & HB\_BR\_FWHM            & float32     & \kms             & FWHM of broad \Hb\ component                               \\
       67 & HB\_BR\_FWHM\_ERR        & float32     & \kms             & Error in HB\_BR\_FWHM                                      \\
\enddata
\end{deluxetable*}

\begin{acknowledgments}
This work is supported by the National Key R\&D Program of China No. 2023YFA1607903, 2022YFF0503402. 
WKR is supported by the China Scholarship Council No. 202306340001. 
HXG is supported by the Future Network Partner Program, CAS, No. 018GJHZ2022029FN, Overseas Center Platform Projects, CAS, No. 178GJHZ2023184MI. The authors acknowledge the use of the iDark High Performance Computing Facility at The University of Tokyo, and associated support services, in the completion of this work.
S.W. acknowledges the support by the National Research Foundation of Korea (NRF) grant funded by the Korean government (MEST) (No. 2019R1A6A1A10073437). J.X.W. acknowledges the support by National Natural Science Foundation of China (grant nos. 12033006, 12192221). 

\end{acknowledgments}

\vspace{5mm}
\facilities{Sloan}

\software{PyQSOFit \citep{2018ascl.soft09008G},
          astropy \citep{astropy:2022},
          pandas \citep{reback2020pandas},
          scipy \citep{2020SciPy-NMeth},
          numpy \citep{harris2020array},
          }

\bibliography{PaperScript}{}
\bibliographystyle{aasjournal}

\end{document}